\documentclass[notitlepage,amsmath,amssymb,aps,reprint,superscriptaddress,footinbib,longbibliography,floatfix]{revtex4-2} 
\usepackage{graphicx}
\usepackage{dcolumn}
\usepackage{bm}
\usepackage{braket}
\usepackage{amsmath}
\usepackage{bm}
\usepackage{amssymb}
\usepackage{setspace}
\usepackage[euler]{textgreek}
\allowdisplaybreaks
\usepackage{multirow}
\usepackage{comment}
\usepackage{relsize}
\usepackage{xr-hyper}
\usepackage{hyperref}
\usepackage{siunitx}
\usepackage{mathtools}
\usepackage{booktabs}
\AtBeginDocument{
\heavyrulewidth=.08em
\lightrulewidth=.05em
\cmidrulewidth=.03em
\belowrulesep=.65ex
\belowbottomsep=0pt
\aboverulesep=.4ex
\abovetopsep=0pt
\cmidrulesep=\doublerulesep
\cmidrulekern=.5em
\defaultaddspace=.5em
}
\hypersetup{
    colorlinks=true,
    linkcolor=blue,
    filecolor=magenta,      
    urlcolor=magenta,
    citecolor=teal
}
\usepackage[x11names]{xcolor}

\newcommand{\ron}{r_{\text{on}}}
\newcommand{\roff}{r_{\text{off}}}

\newcommand{\pe}{p_\mathrm{e}}
\newcommand{\plost}{p_\mathrm{lost}}

\newcommand{\Nmax}{N_{\text{max}}}

\newcommand{\mfa}{\mathfrak{a}} 
\newcommand{\mfb}{\mathfrak{b}} 

\newcommand{\equal}{\!=\!}

\newcommand{\tauG}{\tau_{\!\scriptscriptstyle \mathcal{G}}}
\newcommand{\OmG}{\Omega_{\scriptscriptstyle \mathcal{G}}}

\newcommand{\chill}{\chi^{\scriptscriptstyle (2)} }

\makeatletter
\newcommand*{\addFileDependency}[1]{
  \typeout{(#1)}
  \@addtofilelist{#1}
  \IfFileExists{#1}{}{\typeout{No file #1.}}
}
\makeatother

\usepackage{lineno}

\begin{document}
\title{Zero-Added-Loss Entangled Photon Multiplexing\\ for Ground- and Space-Based Quantum Networks}
\author{Kevin C. Chen}
\affiliation{Research Laboratory of Electronics, Massachusetts Institute of Technology, Cambridge, MA 02139, USA}
\affiliation{Department of Electrical Engineering and Computer Science, Massachusetts Institute of Technology, Cambridge, MA 02139, USA}
\author{Prajit Dhara}
\affiliation{Wyant College of Optical Sciences, The University of Arizona, Tucson, AZ 85721}
\affiliation{NSF-ERC Center for Quantum Networks, The University of Arizona, Tucson, AZ 85721}
\author{Mikkel Heuck}
\affiliation{Department of Electrical and Photonics Engineering, Technical University of Denmark, 2800 Kgs. Lyngby, Denmark}
\author{Yuan Lee}
\affiliation{Department of Electrical Engineering and Computer Science, Massachusetts Institute of Technology, Cambridge, MA 02139, USA}
\author{Wenhan Dai}
\affiliation{Department of Electrical Engineering and Computer Science, Massachusetts Institute of Technology, Cambridge, MA 02139, USA}
\affiliation{College of Information and Computer Science, University of Massachusetts, Amherst, Massachusetts 01003, USA}
\author{Saikat Guha}
\affiliation{Wyant College of Optical Sciences, The University of Arizona, Tucson, AZ 85721}
\affiliation{NSF-ERC Center for Quantum Networks, The University of Arizona, Tucson, AZ 85721}
\author{Dirk Englund}
\email{englund@mit.edu}
\affiliation{Research Laboratory of Electronics, Massachusetts Institute of Technology, Cambridge, MA 02139, USA}
\affiliation{Department of Electrical Engineering and Computer Science, Massachusetts Institute of Technology, Cambridge, MA 02139, USA}
\affiliation{Brookhaven National Laboratory, Upton, NY 11973, USA}

\begin{abstract}
We propose a scheme for optical entanglement distribution in quantum networks based on a quasi-deterministic entangled photon pair source. By combining heralded photonic Bell pair generation with spectral mode conversion to interface with quantum memories, the scheme eliminates switching losses due to multiplexing \textit{in the source}. We analyze this `zero-added-loss multiplexing' (ZALM) Bell pair source for the particularly challenging problem of long-baseline entanglement distribution via satellites and ground-based memories, where it unlocks additional advantages: (i) the substantially higher channel efficiency $\eta$ of \textit{downlinks} vs. \textit{uplinks} with realistic adaptive optics, and (ii) photon loss occurring \textit{before} interaction with the quantum memory -- i.e., Alice and Bob receiving rather than transmitting -- improve entanglement generation rate scaling by $\mathcal{O}(\sqrt{\eta})$. Based on numerical analyses, we estimate our protocol to achieve $>10~$ebit/s at memory multiplexing of $10^2$ spin qubits for ground distance $>10^2~$km, with the spin-spin Bell state fidelity exceeding 99$\%$. Our architecture presents a blueprint for realizing global-scale quantum networks in the near-term.
\end{abstract}
\maketitle

\section{Introduction} 
\label{sec:intro}

Entanglement distribution across distant nodes is fundamental to constructing quantum networks~\cite{Wehner_2018}. However, despite recent progress via optical fiber links~\cite{Pompili_2021,Bhaskar_2020,Ruf_2021}, scaling quantum networks to global reach remains a formidable challenge. One approach to increasing the entanglement rate over low efficiency $(\eta\ll 1)$ elementary links is to use a deterministic Bell-state source at Charlie (C) midway between quantum repeaters (QRs) Alice (A) and Bob (B) in lieu of an entanglement swap, i.e. the midpoint source architecture. If A (B) post-selects on events where a photon passed at least path length $\overline{\text{AC}}$ ($\overline{\text{BC}}$), in the regime where the QRs are memory-limited, the average entanglement rate $\bar{\Gamma}$ improves to ${\propto\sqrt{\eta}}$ compared to ${\propto\eta}$ in typical protocols where C performs local  Bell state measurements (BSM)~\cite{Jones_2016}. Despite the increase in QR complexity, this ``midpoint source'' scheme enables an increase in $\bar{\Gamma}$ by $\mathcal{O}(1/\sqrt{\eta})$ and has motivated research efforts to produce the required entangled photon pair sources, which should be near-deterministic for the advantage to persist. Thus far, leading efforts are based on cascaded atomic sources~\cite{Parniak_2017,Lipka_2021} or spontaneous parametric down conversion (SPDC) sources. While the latter has demonstrated heralded production of Bell pairs out of a single~\cite{Zhang_2008,Barz_2010,Fldzhyan2021-pw,Stanisic2017-we} or a pair of SPDC sources~\cite{Mower_2011,Dhara_2021}, the existing approaches still suffer in either efficiency or fidelity required for near-term quantum networks.

\begin{figure*}[t]
    \centering
    \includegraphics[width=0.8\textwidth]{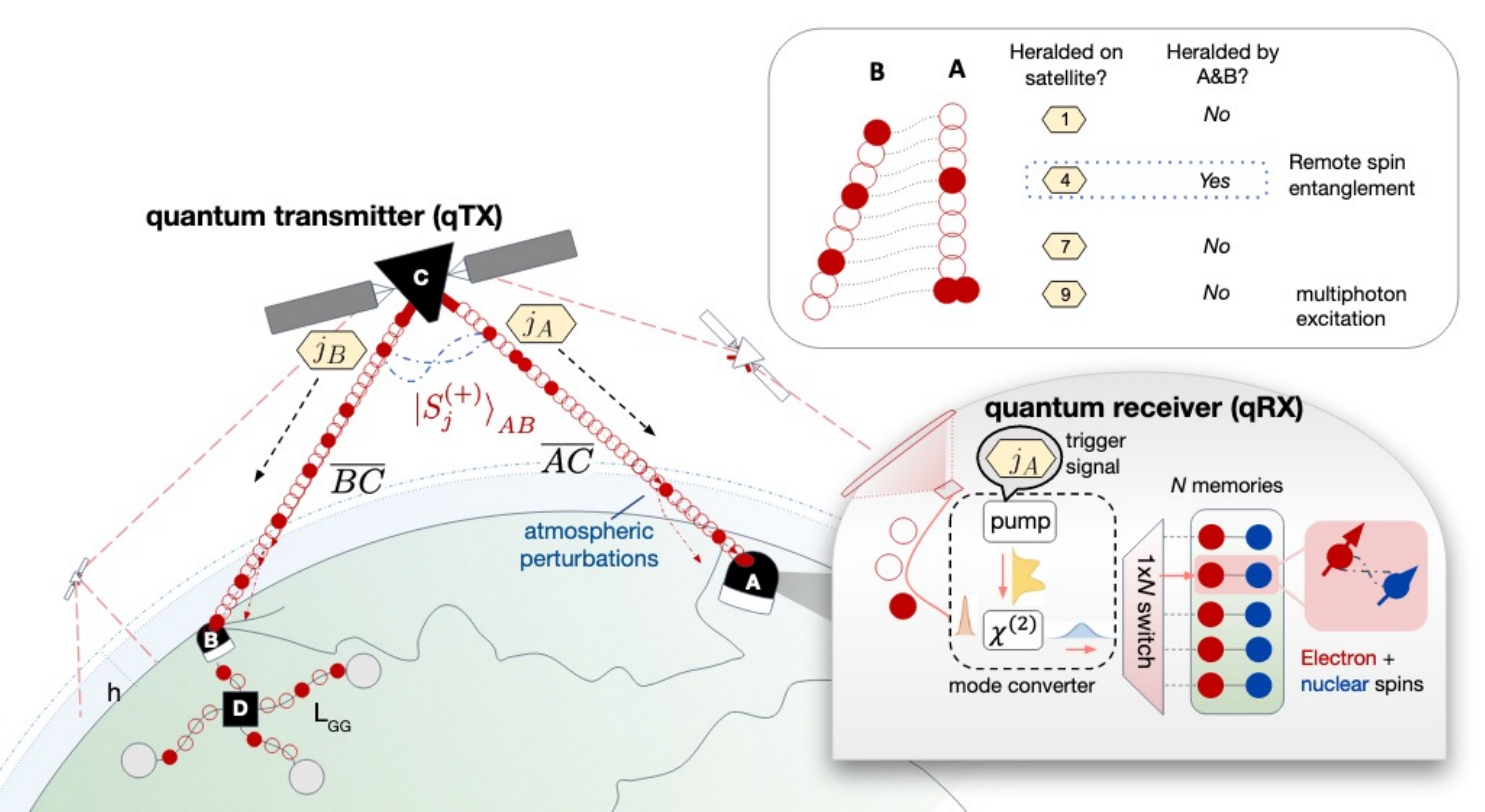}
    \caption{Protocol overview --- A global-scale quantum network composed of flying satellites as the quantum transmitters (qTX) and terrestrial stations as the quantum receivers (qRX). The qTX (C) emits heralded photonic Bell pairs $|S_j^{(+)}\rangle_{AB}$ to A and B, followed by subsequent spectral mode conversion. Filled (open) circles represent successfully  \textit{heralded} (lost) Bell pairs. The two bunched photons signify multi-photon emission due to BPS's imperfection. $|S_j^{(+)}\rangle_{AB}$ is accompanied by a classical signal $j_{A,B}$ containing frequency information. A 1x$N$ switch then routes it to an arbitrary spin memory composed of electron-nuclear spins. The qTX containing quasi-deterministic heralded BPS can also be implemented in ground-only quantum networks, as shown by node $D$ connected to other network stations, separated by ground distance $L_\text{GG}$ (considered in Appendix~\ref{app_ground_networks}).}
    \label{fig:architecture}
\end{figure*}

In this Article, we propose a quasi-deterministic Bell pair source (BPS) that eliminates compounding switching loss of previously proposed spatially multiplexed~\cite{Dhara_2021} or purely temporally multiplexed~\cite{Mower_2011} heralded BPS. Namely, this ``zero-added-loss multiplexing'' (ZALM) BPS leverages the large SPDC phase-matching bandwidth to achieve high transmission rate via \textit{spectral} multiplexing.

 A high bandwidth BPS is crucial to general long-distance entanglement distribution to compensate for channel losses. In this proposal, we consider one specific application of utilizing space-to-ground optical links to establish ground-to-ground entanglement, though such a quasi-deterministic BPS would benefit any general quantum network setting. Reaching the \textit{global} scale, however, calls for space-based configurations, one example of which is shown in Fig.~\ref{fig:architecture}, where the midpoint source is a satellite sending entangled photons to two distant ground stations. This satellite-mediated entanglement distribution scheme has to contend with extreme channel conditions. Here we consider three primary factors that contribute to our particular setup: large transmission loss, channel instability, and heralding latency.

Two canonical choices for satellite-based optical transmission links are the \emph{downlink} and \emph{uplink} configurations. Despite the relative ease of just having an interferometric system for BSM in space, the uplink suffers from pointing instability due to the ``shower-curtain effect''~\cite{Dror_1998}, requiring two-way adaptive optics yet to be demonstrated, making the downlink configuration more efficient~\cite{Pirandola_2021}. Therefore, we limit our investigation to a two-photon downlink architecture.

A single-rail (heralding on one photon)~\cite{Pompili_2021,Humphreys_2018} encoded protocol could increase $\bar{\Gamma}\propto\eta$ over a dual-rail (heralding on two detected photons)~\cite{Barrett_2005} scheme with $\bar{\Gamma}\propto\eta^2$. However, it is presently unclear whether the optical carrier-level phase tracking required in the single-photon scheme is compatible with space-to-ground links.

Specifically, we assume: (1) the availability of memory-multiplexed ($k$-qubits) QRs on the ground -- but not on the satellite; (2) satellite-to-ground (SG) downlinks; (3) SG transmission in the telecommunications C-band via frequency conversion for compatibility with space-qualified photonics~\cite{Bozovich_2021}; and (4) two-photon-heralding~\cite{Pompili_2021,Humphreys_2018}.

This Article is organized as follows. Section~\ref{sec:overview} presents an overview of the satellite-assisted entanglement distribution architecture, one specific example of global-scale quantum network utilizing the ZALM BPS. Section~\ref{sec:quantum_transmitter} discusses the quantum state description of a heralded entangled state and evaluates the boosted BPS emission rate with spectral multiplexing. Section~\ref{sec:quantum_receiver} then describes the two critical components in the ground-based quantum receiver: (1) a mode converter for temporal-spectral conversion via a sum-frequency generation process; (2) a cavity-based spin-photon interface based on a hybrid photonic integrated circuit. We then evaluate the ground-to-ground spin-spin entanglement state fidelity and efficiency in Sections~\ref{sec:fidelity}-\ref{sec:rate}, accounting for imperfections in the ZALM BPS and the ground-based quantum receivers. Additionally, we discuss the trade-off between entanglement fidelity and generation rate due to unheralded photon loss in the channels. Finally, Section~\ref{sec:conclusion} concludes the paper and offers thoughts on alternative approaches for implementing the spectrally multiplexed BPS. Importantly, it addresses engineering challenges that still need to be overcome in order to realize the proposed architecture.

\section{Entanglement Distribution Architecture Overview} \label{sec:overview}

Improving efficiency and fidelity of quantum links for scalable quantum networks beyond recent satellite-based demonstrations~\cite{Liao_2017,Yin_2020} likely requires heralded entanglement of quantum memories. Here, we propose an architecture integrating the merits of satellite-based channels and spin-photon interfaces containing diamond color centers. On the satellite, C as a multiplexed quantum transmitter (qTX) emits heralded polarization-encoded photonic Bell states $|S_j^{(+)}\rangle_{AB}=(\ket{H_AV_B}+\ket{V_AH_B})/\sqrt{2}=(\ket{1,0;0,1}+\ket{0,1;1,0})/\sqrt{2}$ (polarization-Fock representation~\cite{Dhara_2021}), accompanied by a classical heralding message encoding its frequency information, $j_{A,B}$. Idler photons (of same polarization unknown to us) of a pair of pulse-pumped SPDC sources are interfered and detected in \textit{wavelength demultiplexed} channels to herald Bell pairs in the signal photons. The detail will be covered in Section~\ref{sec:quantum_receiver}.

Each photon of the Bell pair described by $|S_j^{(+)}\rangle_{AB}$ then travels to its respective terrestrial quantum receiver (qRX), A and B (detailed in Section~\ref{sec:quantum_receiver}).  At each receiver, a mode-converter (MC)~\cite{Heuck_2020_PRA,Heuck_2020_PRL} converts both the frequency and the spectral bandwidth of the photon to match those of the quantum memories, crucial for efficient cavity-based spin-photon interaction~\cite{Duan_2004}.  Finally, successful photon detection at both A and B completes teleportation, i.e.\ A and B sharing spin-spin entanglement upon heralding on the same Bell state's photons. 

 The analysis and discussion of our proposal in the rest of manuscript is tailored towards studying the performance and challenges of a single quantum link; however the utility of a ZALM link is applicable to terrestrial fiber-based quantum links. Multi-hop linear chain of quantum repeaters (where the channel between A and B is subdivided into smaller segments) with ZALM architecture based links for generation of entanglement and local entanglement swapping between memories (in a repeater station) is a natural extension of our study, as is a regular grid network (on a well defined `lattice') of repeaters which can support multiple entanglement generation paths. We reserve the analysis of these advanced network geometries for future work.
 
\section{Quantum transmitter} \label{sec:quantum_transmitter}
\subsection{Photonic sources of high-quality dual-rail entangled qubits} 

Generation of high-quality photonic entangled pairs is an open research challenge, with multiple possible approaches, each having their own set of merits and demerits. We consider SPDC sources pumped with a mode-locked laser, combining previous proposals for quasi-deterministic sources~\cite{Zhang_2008,Barz_2010,Zhong_2018,Dhara_2021,Fldzhyan2021-pw,Stanisic2017-we,Mower_2011, Mower_2013, Heuck_2018, Zhao_2020}. More specifically, our BPS builds on the proposal of Ref.~\cite{Dhara_2021} of interfering photons of a pair of SPDC sources~\cite{Kwiat_1999,Bouwmeester_1997,Kok_2000} to herald Bell pair production. Prior proposals necessitate spatial multiplexing to boost the Bell pair generation rate. However, this requires optical switches in a tree-configuration whose number of layers grows exponentially with the number of spatial multiplexing modes. Given finite loss per switch, the compounded optical loss through the switch array quickly renders the Bell pair generation inefficient. For example, the multiplexing requirements are very demanding ($\sim 10^7$ sources running in parallel) to achieve quasi-deterministic state generation. Considering 0.5~dB loss per switch, $\sim 10^7$ would lead to $10^{-0.5/10\cdot \log_2(10^7)}\sim 7\%$ efficiency.

\begin{figure}[t]
    \centering
    \includegraphics[width=0.5\textwidth]{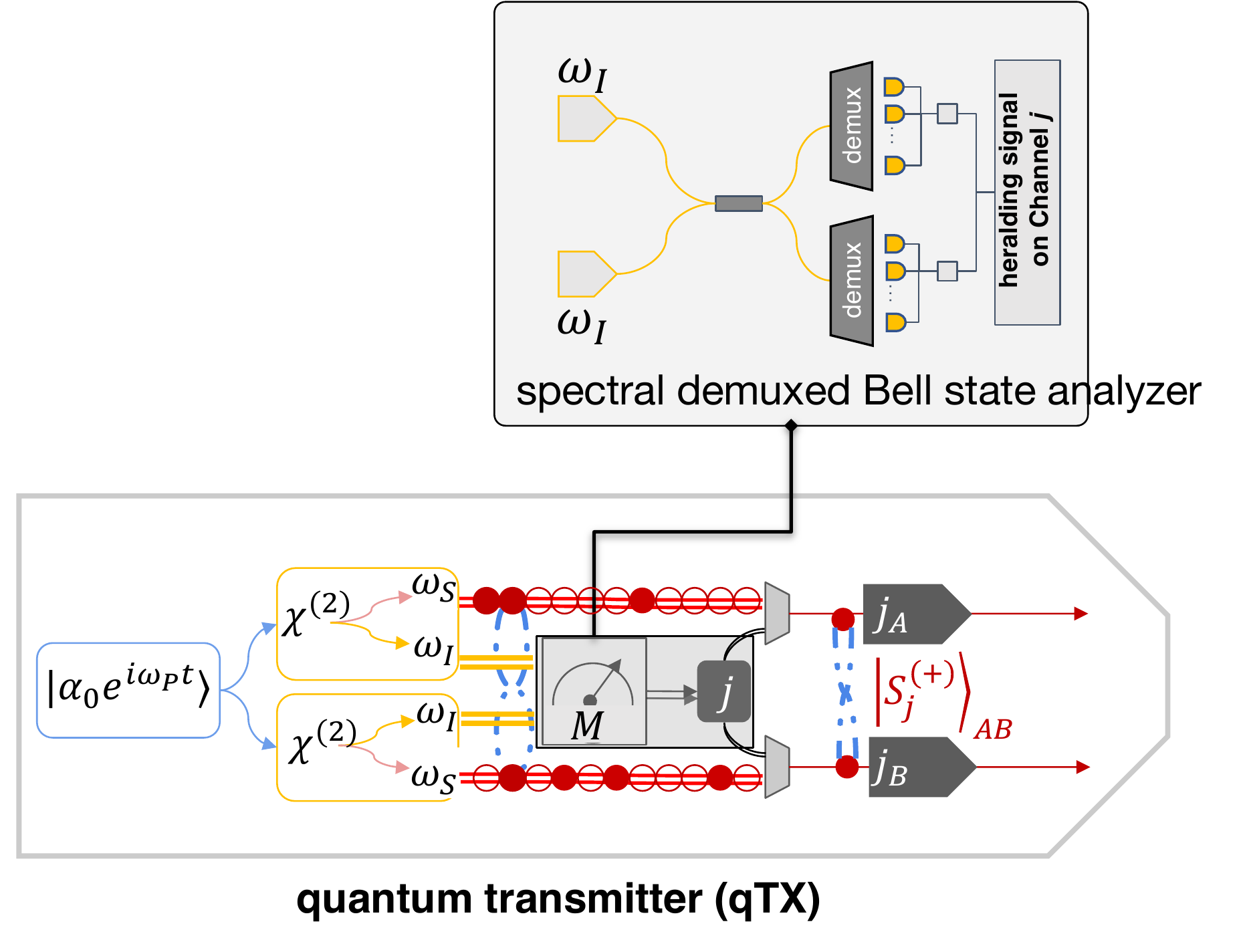}
    \caption{ Implementation of qTX. -- A multiplexed BPS comprises a pair of pulse-pumped SPDC sources. A spectral demultiplexed BSM heralds the Bell pair creation and sends out classical messages encoding its frequency information.}
    \label{fig:qTX}
\end{figure}

In our current proposal, we leverage the large phase-matching bandwidth $\sigma_\text{pm}=10~$THz for spectral multiplexing. After the beam splitter interaction of the intermediate BSM, we demultiplex the output beams into dense wavelength division multiplexing channels (DWDM) and perform single photon detection on each channel separately  (Bell state analyzer shown in Fig.~\ref{fig:qTX}). Each DWDM channel is spaced apart by $\sigma_{\text{ch}}=12.5~$GHz over $\sigma_\text{pm}$ in the C-band.  If detection (with the correct click patterns; see Appendix~\ref{app_down_conv}) occurs in the same frequency channels for the two DWDMs, then the BPS has heralded production of a photonic Bell state.

First, we consider the quantum state of a single down-conversion process in a $\chi^{(2)}$ medium~\cite{Grice1997-rv,Grice2001-fk,Mower_2013},
\begin{align} \label{eqn:DC_state} 
\ket{\psi} &= c_0 \ket{\textbf{0}}+c_1 \int J(\omega_S,\omega_I)\hat{a}_S^{\dagger} (\omega_S) \hat{a}_I^{\dagger} (\omega_I)\ket{\textbf{0}} \nonumber\\
&\quad + c_2 \int J(\omega_{S},\omega_I)\hat{a}_S^{\dagger} (\omega_S)\hat{a}_S^{\dagger}  (\omega_I)\nonumber\\
&\quad\quad \quad \; \cdot J(\omega'_S,\omega'_I)\hat{a}_S^{\dagger} (\omega'_S)\hat{a}_I^{\dagger} (\omega'_I)  \ket{\textbf{0}}
\end{align}
 up to two-photon contributions, where $\ket{\textbf{0}}$ is signal-idler mode in the vacuum state, and $\hat{a}^\dagger({\omega_k})$ is the creation operator at frequency $\omega_k$ for the signal ($k=S$) or idler $(I). $. $J(\omega_S,\omega_I)$ represents the joint spectral amplitude function (Appendix~\ref{app_down_conv}).

The constituent terms of Eq.(\ref{eqn:DC_state}) are the vacuum ($\ket{0,0}$), single photon entanglement (with $J(\omega_S,\omega_I) $ and terms with a second order contribution. The latter are detrimental to entanglement distribution protocols as they lie outside the dual-rail photonic qubit Hilbert space. With photon loss, these terms yield false click patterns and limit the fidelity of the distributed entangled pairs~\cite{Dhara_2021,Krovi2016-ma}. Each SPDC source in Fig.~\ref{fig:qTX} comprises two down-conversion processes \footnote{Although separate SPDC sources have been shown here, the same could be achieved by a single source and a spectrally demultiplexed Franson interferometer for time-basis entanglement swap.}. The qTX performs BSM in each DWDM channel by interfering the idler photons from both SPDC sources. We take $J(\omega_S,\omega_I)$ to exhibit frequency anti-correlation between signal-idler photons, such that a demultiplexed detection heralds `which-frequency' information about the Bell pair. Based on the analysis of Ref.~\cite{Dhara_2021} and its extension to the current proposal (see Appendix~\ref{app_down_conv} for a detailed derivation), the spectrally-synchronized
BSM that yields one of the desirable photon click patterns (say on channel $j$) heralds an entangled state with
a spectral description given as,
\vspace{-4em}
\begingroup\makeatletter\def\f@size{9}\check@mathfonts
\def\maketag@@@#1{\hbox{\m@th\large\normalfont#1}}%
\begin{widetext}
     \begin{align}
	\ket{S^{(\pm)}_j}\propto 	 &\int_{\vec{\Omega}_j} d\bar{\omega} \Biggl[  \biggl( J(\omega_{A_1},\omega_{A'_1})  \, J (\omega_{B_2'},\omega_{B_2})\, \hat{a}^\dagger_{A_1} (\omega_{A_1}) \, \hat{a}^\dagger_{B_2} (\omega_{B_2})
		+(-1)^{m_1} J(\omega_{A_2},\omega_{A_2'})  \, J (\omega_{B_1'},\omega_{B_1})\, \hat{a}^\dagger_{A_2} (\omega_{A_2})\, \hat{a}^\dagger_{B_1} (\omega_{B_1})\biggr)\nonumber \\ 
		&+(-1)^{m_2}\biggl( J(\omega_{A_1},\omega_{ A_1'}) \,J(\omega_{A_2},\omega_{A_2'})  \, \hat{a}^\dagger_{A_1}(\omega_{A_1})\, \hat{a}^\dagger_{A_2}(\omega_{A_2})
		+  (-1)^{m_1} J (\omega_{B_2'},\omega_{B_2}) \,J (\omega_{B_1'},\omega_{B_1})\, \hat{a}^\dagger_{B_1} (\omega_{B_1})\, \hat{a}^\dagger_{B_2} (\omega_{B_2})\biggr)
		\Biggr] \nonumber\\
		&\times \ket{0}_{A_1}  \ket{0}_{A_2}  \ket{0}_{B_1} \ket{0}_{B_2}.
		\label{eqn:bigstatem}
	\end{align}   
\end{widetext}
\endgroup
Here, the modes $A_1,A_2 \,(B_1,B_2)$ correspond to the qubits transmitted to A (B), and $\int_{\vec{\Omega}_j} d\bar{\omega}$ represents integration over the $j$-th detection channel's spectral bandwidth $\vec{\Omega}_j$. Subscript 1 (2) represents the polarization mode $H$ ($V$). Parity bits $m_1$ and $m_2$ depend on the detection pattern and determine the distributed entangled state (Appendix~\ref{app_down_conv}).

The first two terms in Eq.~(\ref{eqn:bigstatem}) represent having one photon each in A's and B's frequency correlated channels, corresponding to the quantum state equivalent to $(\ket{1,0}_A\ket{0,1}_B\pm\ket{0,1}_A\ket{1,0}_B)/\sqrt{2}=(\ket{H}_A\ket{V}_B\pm\ket{V}_A\ket{H}_B)/\sqrt{2}\equiv\ket{\Psi^\pm}$ in the spectral-mode basis. The remaining terms of Eq.~(\ref{eqn:bigstatem}) represent a state in which A (B) receives both photons and B (A) receives none, i.e. it is equivalent to a term of the form, $(\ket{1,1}_A\ket{0,0}_B\pm\ket{0,0}_A\ket{1,1}_B)/\sqrt{2}$ in the polarization-Fock basis, contributing to the heralded state infidelity.  We note that leakage photons stemming from the second-order term in Eq.~\ref{eqn:DC_state} with non-degenerate frequencies (i.e. $\omega_s\neq\omega_s'$) would not introduce additional errors in spin-spin entanglement, as discussed later in Section~\ref{subsec:mode_converter}.


\subsection{Average Bell pair generation rate} \label{subsec:source_rate}

In calculating the average emission rate of the ZALM BPS, we abstract all the parameters of the transmission and collection optics into a single channel loss parameter $\sqrt{\eta}$ for each satellite-ground link; assuming suitable levels of timing synchronicity, adaptive optics, pointing and tracking, Doppler compensation and beam forming for the architecture design.

\begin{figure}[t]
    \centering
    \includegraphics[width=0.5\textwidth]{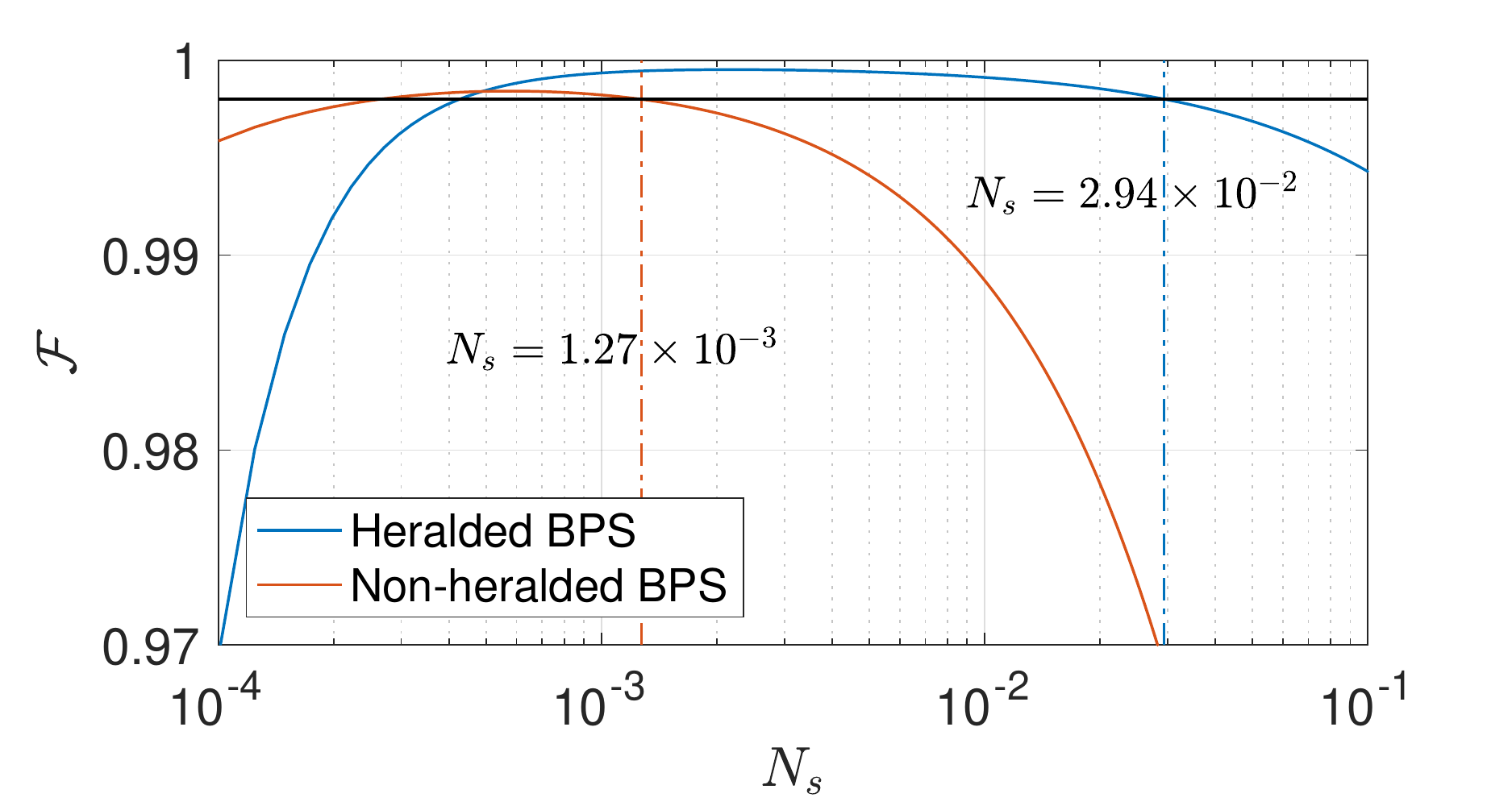}
    \caption{The spin-spin Bell state fidelity $\mathcal{F}$ as a function of the mean photon number $N_s$ for both the heralded BPS (blue solid) and the non-heralded free-running narrowband BPS (orange solid). The horizontal black solid line indicates the fixed fidelity $\mathcal{F}=0.998$, which is chosen based on particular parameter values considered (see main text). Both the blue and orange dashed lines indicate their intersections at the respective $N_s$ values for the heralded and non-heralded BPS.}
    \label{fig:BPS_find_Ns}
\end{figure}

The essence of the heralded BPS relies on leveraging the entirety of the phase matching bandwidth of the SPDC, which we assume to be $\sigma_{\text{pm}}=10~$THz. To compute the BPS emission rate requires knowing the density matrix of the heralded BPS' output state $\rho_\mathrm{BPS}$, which depends on the mean photon number per mode $N_s$ that is effectively dictated by the power of the pump field. Given $\rho_\mathrm{BPS}$ and accounting for imperfections in the quantum receiver, we can evaluate the resultant spin-spin entangled state and its fidelity to the ideal Bell state, $\mathcal{F}$, after heralded teleportation of the photonic qubits. More details on calculations of $\mathcal{F}$ will be covered in Section~\ref{sec:quantum_receiver}. Figure~\ref{fig:BPS_find_Ns} shows $\mathcal{F}$ as a function of $N_s$ for both the heralded BPS and the non-heralded free-running narrowband SPDC source. For the former, as $N_s$ increases, the effect of loss is suppressed and hence the fidelity increases. For the latter, however, fidelity increases from having higher Bell pair contribution relative to the other order terms. After reaching an optimum, the fidelity begins to drop with increasing $N_s$ as a result of multi-photon events degrading the photonic Bell state fidelity. Note that in general the generation rate can be improved by increasing $N_s$, at the cost of lowering the heralded photonic Bell state fidelity (thereby the spin-spin Bell state fidelity). For an instance, for the ZALM BPS, increasing $N_s$ to $8\times 10^{-2}$ still maintains a fidelity $\mathcal{F}\geq 0.995$. On the other hand, with a free-running SPDC, the fidelity drops off much more quickly due to higher order terms that cannot be eliminated via heralding.

To fairly compare the performance of the two, we fix the spin-spin Bell state fidelity still attainable by the non-heralded BPS at $\mathcal{F}=0.998$ and derive a corresponding a mean photon number of $N_s=2.94\times 10^{-2}$ for the heralded ZALM BPS. Given $N_s$, we numerically compute the density matrix of the heralded photonic state in the Fock basis for each spectral mode, and consequently calculate the probability of generating a heralded Bell pair $p_{\mathrm{gen}}$ by taking a partial trace over the Bell basis states. We find $p_{\mathrm{gen}}=3.6\times 10^{-4}$ per spectral mode. Now, accounting for all the spectral modes, we can greatly boost the generation probability. Specifically, assuming a DWDM channel bandwidth of 12.5~GHz, the number of spectral-multiplexed modes is simply $N_{\text{modes}}=\sigma_{\text{pm}}/\gamma=10~$THz$/$12.5~GHz$=800$. The probability of successfully heralding a photonic Bell pair per \textit{pulse} across the entirety of the phase-matching bandwidth $\sigma_\text{pm}$ is $p_{\text{ZALM}}=1-(1-p_{\text{gen}})^{N_{\text{modes}}}=2.5\times 10^{-1}$. If we consider a state of the art mode-locked laser with a pulse generation repetition rate of $\sigma_{\text{rep}}=30~$GHz~\cite{Hu_2022}, the average Bell pair emission rate is $p_{\text{ZALM}}\times \sigma_{\text{rep}}\,\approx 7.45$~GHz.

In contrast, for a narrowband-filtered free-running SPDC source (non-heralded BPS), we find the required mean photon number mode of $N_{s,\text{SPDC}}=1.27\times 10^{-3}$, to suppress higher order terms and match the fidelity of the heralded spin-spin state ($\mathcal{F}=0.998$) with the heralded ZALM BPS. We again compute its density matrix and find the probability of generating a Bell pair to be $p_{\text{gen,SPDC}}=2.5\times 10^{-3}$. Since we assume a narrowband-filtered source, the absence of spectral multiplexing means the success probability is the generation probability itself, i.e. $p_{\text{gen,SPDC}}=p_{\text{SPDC}}$. Assuming the same pulsed laser repetition rate of 30~GHz, the average generation rate for the photonic Bell pair is $p_{\text{SPDC}}\times\sigma_\text{rep}=75~$MHz.

\subsection{Requirements on the photon detectors for entanglement swap}

Here we consider the requirements on the single photon detectors in qTX. First, the detector reset time should be short enough such that ideally there are no missed heralding events. Given $N_s$, the average production rate for the signal-idler pair would be $N_s\sigma_{\text{rep}}$. The number of incident photons per channel within the detector's reset time $\tau_r$ would be $\mu=\tau_r N_s \sigma_{\text{rep}}/(4\times N_{\text{modes}})$, where the factor of 4 stems from the need of four detectors for BSM~\cite{Dhara_2021}. Consequently, the probability of detecting $\geq 2$ photons in one channel within the detector's reset time is ${p=1-\exp(-\mu)-\mu\exp(-\mu)}$ following the Poisson distribution. 

We take a detector reset time $\tau_r=1$~ns~\cite{Tarkhov_2008,Annunziata_2010}, resulting in $\mu=1~\text{ns}\times 7.45~\text{MHz}/(4\times 800)=2.3\times 10^{-3}$. The probability of detecting $\geq 2$ photons is then $\sim 2.7\times 10^{-6}\ll 1$, which is sufficiently small to neglect missed detection events due to multiple-photon incidence on the same detector within its reset time.

Second, the detector jitter must be sufficiently small to avoid projecting the output of the ZALM BPS to a mixed state. Otherwise, this would lead to sub-optimal BSM due to spectral distinguishability and an overall reduction in the mode conversion efficiency at the ground stations. To avoid uncertainty in projection within the 12.5~GHz spectral window given by the DWDM channel bandwidth, we need the jitter time to be much less than {$1/12.5~$GHz$~=80~$ps}. Based on Ref.~\cite{Korzh_2020}, we assume the detector jitter time to be close to state of the art at 1~ps. In Appendix~\ref{app_photonic_bell_state}, we also evaluate the impact of having jitter time much larger than $1/12.5~$GHz$\approx 80~$ps on the heralded photonic Bell pair. Namely, having an imperfect measurement would lead to a reduced interference visibility in the BSM, consequently degrading the fidelity of the photonic Bell state. However, with an assumed Gaussian spectral profile per DWDM channel, the JSI almost recovers perfect purity and results in a theoretical maximum visibility of $v=0.996$ (corresponding to a photonic Bell state measurement infidelity of $2\times 10^{-3}$). Further complications concerning mode conversion and spin-photon quantum teleportation with increased jitter time are beyond the scope of this Article and thus tabled for future studies.

\section{Quantum Receiver} \label{sec:quantum_receiver}

\begin{figure}[t]
    \centering
    \includegraphics[width=0.5\textwidth]{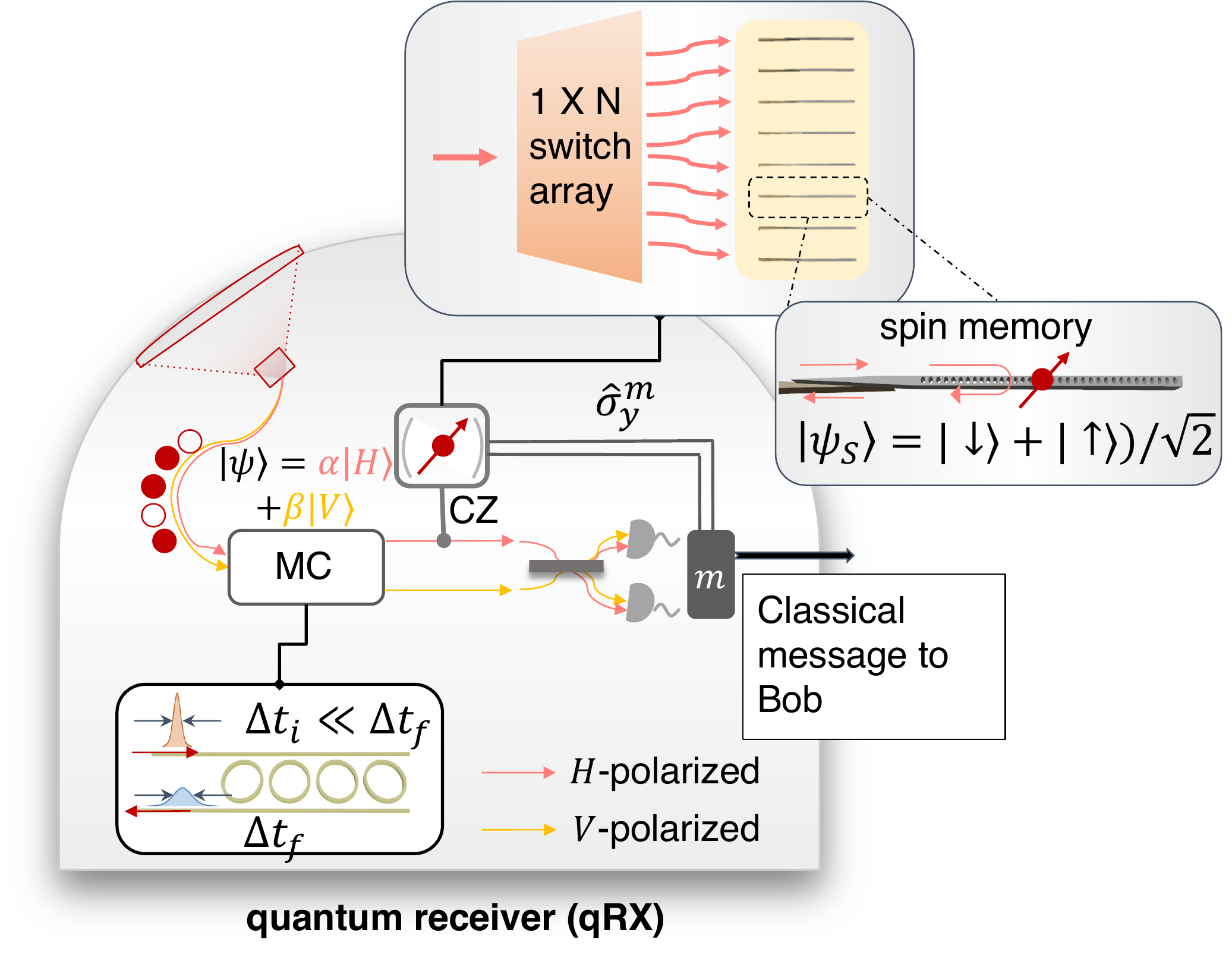}
    \caption{ Implementation of qRX. -- The qRX contains a mode converter (MC) and a $1\times N$ switching array that routes the photon to arbitrary channels in the memory bank, each containing a diamond nanocavity coupled with an optically active electron spin (red).}
    \label{fig:qRX}
\end{figure}

\subsection{Mode conversion for frequency and bandwidth matching} \label{subsec:mode_converter}

For efficient cavity-based spin-photon interaction, the photon must be 1) resonant with the spin's optical transition and, 2) narrow in bandwidth relative to the emitter's optical linewidth~\cite{Duan_2004}. Hence, we consider a ``mode-converter'' (MC) that frequency up-converts from telecom C-band to visible wavelengths and pulse shapes to reduce the photon's spectral bandwidth,  i.e.\ increasing the initial temporal width of the photon $\Delta t_i$ to a final temporal width $\Delta t_f\gg\Delta t_i$.

As shown in Fig.~\ref{fig:qRX}, upon arrival from the satellite, the photon with carrier frequency $\omega_a$ enters a high-Q ring cavity~\cite{Zhang_2017} made from a $\chill$-material, supporting three resonance frequencies: $\omega_a,\omega_b$ (target frequency matching the emitter's transition at 737~nm, see Section~\ref{subsec:spin_photon_interface}) and $\omega_c$ (strong pump's frequency) that satisfy energy conservation: $\omega_a+\omega_c=\omega_b$~\cite{Guo_2016}. Upon receiving the classical message $j_{A,B}$ encoding $|S_j^{(+)}\rangle_{AB}$'s frequency information, the pump field is optimally shaped to convert photons in mode $a$ to mode $b$~\cite{Heuck_2020_PRL,Heuck_2020_PRA}. Since each DWDM channel corresponds to a unique frequency, each ground receiver contains an array of ring resonators whose resonance frequencies are spaced 12.5~GHz apart. To reduce the intensive requirement of having many ring resonators, each resonator may include phase shifters to address multiple DWDM channel frequencies. For example, if each resonator has up to 1~THz of tuning range~\cite{Xue_2016,Wang_2018,Dong_2021,Panuski_2022}, corresponding to 80 DWDM channels, the number of ring resonators needed reduces from $N_{\text{modes}}=800$ to 10.

A subtle feature of the MC is that it acts as an additional \textit{spectral filter}, which removes spurious spectral modes from interacting with the spin memories in the qRX. For example, second-order terms with two photon-pairs per SPDC source as indicated in Eq.~\eqref{eqn:DC_state} may lead to interference between spectrally degenerate photons (for e.g.\ single idler photons at $\omega_I$ from the constituent SPDC sources) \textit{with an accompanying non-degenerate photon} (i.e.\ one of idler modes actually generates a two-pair term with $\omega_I\neq\omega'_I$ );  henceforth we shall term these as ``leakage photons''. With perfect detection efficiency, these events would be immediately flagged based on the detection pattern (since the pattern would deviate from the ideal spectral mode-synchronized detection). However, in the practical case of non-unity detection efficiency, these events may not be discernible from a scenario in which only first-order photon-pairs from the two SPDC sources interfere (i.e.\ idler photon at $\omega'_I$ is lost). Given a particular BSM with the correct click pattern, the qTX sends the known information $j_{A,B}$ about the heralded Bell pair's frequency. Regardless of the presence of leakage photons, say at  $\omega_S'$, the MC would only up-convert $\omega_S$ photons specified by $j_{A,B}$. In this sense, the final spin-spin Bell state is partially post-selected spectrally. We note that the edge case of having second-order terms with spectrally degenerate idler photons (and therefore signal by frequency correlation) is indeed a non-correctable error, but is accounted for in the fidelity calculations presented in Section~\ref{sec:fidelity} later.

We use the method developed in Ref.~\cite{Heuck_2020_PRA} to realize a beam splitter Hamiltonian (in the frequency domain) between modes $a$ and $b$ via sum-frequency-generation (SFG) between the pump and mode $a$, though other proposed methods are equally viable~\cite{Fejer_1994,Brecht_2015,Salem_2008,Myilswamy_2020,Yu_2021}. The Hamiltonian describing the cavity modes is 
\begin{align}
    \hat{H} = \hbar\chi_{\scriptscriptstyle {\rm{SFG}}} \big( \hat{a}\hat{b}^\dagger\hat{c} + \hat{a}^\dagger\hat{b}\hat{c}^\dagger \big) + \sum_{q} i\hbar \sqrt{\kappa_{q,\rm{w}}} \big( \hat{q}^\dagger \hat{w}_q  - \hat{w}_q^\dagger  \hat{q}\big), 
\end{align} 
where $q=\{a,b,c\}$, $\hat{w}_q$ is the annihilation operator of the input (i.e. waveguide) mode interacting with cavity mode $q$, and $\kappa_{q,\rm{w}}$ is the cavity-waveguide coupling rate. $\chi_{\scriptscriptstyle {\rm{SFG}}}$ is the SFG coefficient. If mode $c$ contains the strong pump mode, it may be treated classically, and the first term in the Hamiltonian has the beam splitter form~\cite{Heuck_2020_PRA}
\begin{align}\label{H beam splitter}
    \hat{H}_{\scriptscriptstyle {\rm{BS}}}  = \hbar \Big( \Lambda^{\!*}(t) \hat{a}^\dagger\hat{b} + \Lambda(t)\hat{a}\hat{b}^\dagger \Big) ,
\end{align} 
where $\Lambda(t) \equal \chi_{\scriptscriptstyle {\rm{SFG}}} \langle \hat{c}(t)\rangle \equal \chi_{\scriptscriptstyle {\rm{SFG}}} \sqrt{n_c(t)}$ with $n_c$ being the number of pump photons in the control mode, $c$. Taking the quantum state of the cavity to be
\begin{align}
    \ket{\Psi_{\rm{cav}}} \equiv [\psi_a(t) \hat{a}^\dagger  +  \psi_b(t) \hat{b}^\dagger] \ket{0}_a\ket{0}_b,
    \label{eq:qs_cavity}
\end{align} 
we can derive the equations of motion from the Schrodinger equations, which are detailed in Appendix~\ref{app freq conv}. The input state may be expressed in a time-bin basis as~\cite{Heuck_2020_PRA} 
\begin{align}
    \ket{\Psi_{\rm{in}}} \equiv  \int d t \xi_{a,\rm{i}} \hat{w}_a^\dagger(t) \ket{\boldsymbol{0}}_t,
    \label{eq:qs_input}
\end{align} 
with $\ket{\boldsymbol{0}}_t$ representing the temporal multi-mode vacuum state $\hat{w}_a^\dagger (t)$ populates the time-bin indexed by $t$ with one photon. The output state is 
\begin{align}
    \ket{\Psi_{\rm{out}}} \equiv   \int d t \xi_{a,\rm{o}}(t) \hat{w}_a^\dagger(t) \ket{\boldsymbol{0}}_t +  \int d t \xi_{b,\rm{o}} (t) \hat{w}_b^\dagger(t) \ket{\boldsymbol{0}}_t,
    \label{eq:qs_output}
\end{align}
where $\xi_{a,\rm{o}},\xi_{b,\rm{o}}$ are two Schrodinger coefficients solvable by the equations of motion (see Appendix~\ref{app freq conv}). We assume that a maximum conversion efficiency is achieved when $\xi_{a,\rm{o}}(t) \equal 0$ by solving for an optimally shaped control pulse described by $\Lambda(t)$.

As an example, let us consider Gaussian input pulses
\begin{align}
\xi_{a,\rm{i}}(t) &=  \!\sqrt{\frac{2}{\tauG}} \!\left(\frac{\text{ln}(2)}{\pi}\right)^{\!\!\frac{1}{4}} \!\!\exp\!\left(\!-2\text{ln}(2)\frac{(t-T_{\rm{in}})^2}{\tauG^2} \right), 
\label{eq:temp_Gaussian}
\end{align} 
where $|\xi_{a,\rm{i}}(t)|^2$ has a full temporal width at half maximum (FWHM) of $\tauG$, spectral width of $\OmG\equal 4\text{ln}(2)/\tauG$, and integrates to 1 (over the infinite interval from $-\infty$ to $\infty$).

\begin{figure}[t]
    \centering
    \includegraphics[width=0.4\textwidth]{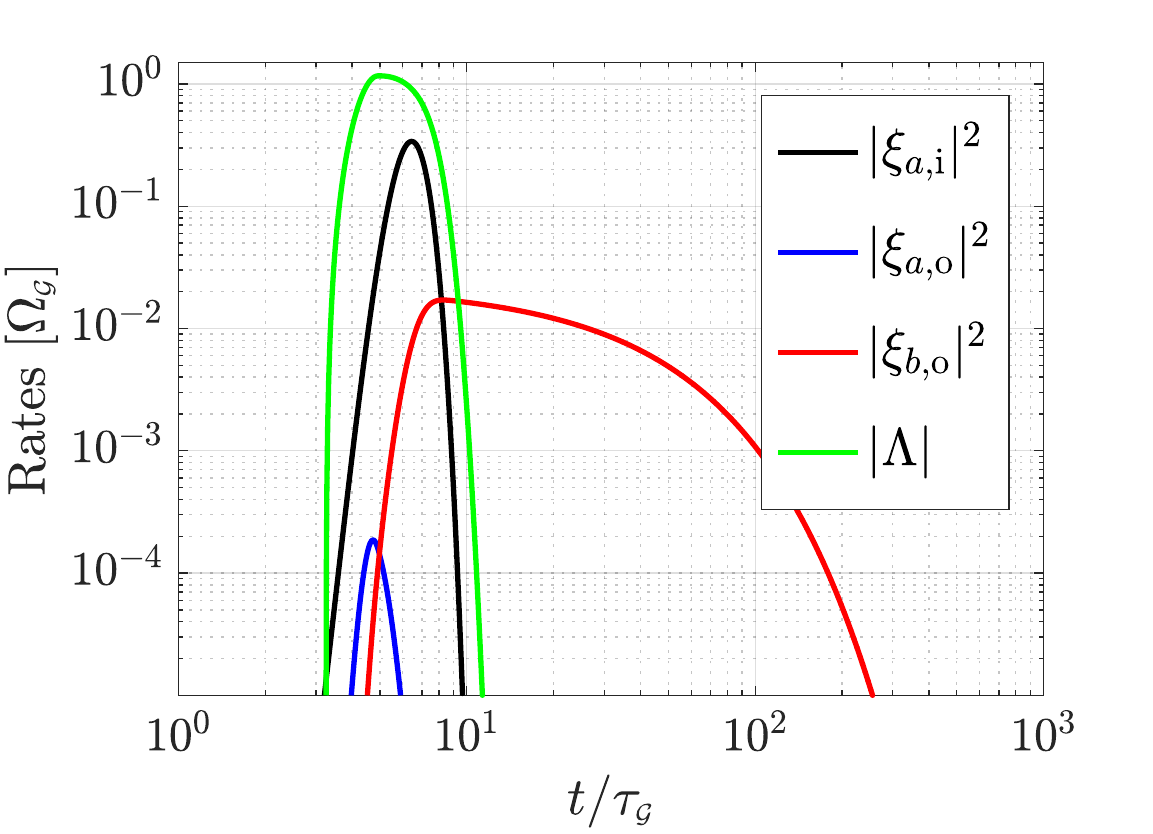}
    \caption{Simulation results for frequency conversion and bandwidth narrowing. $\xi_{a,\rm{i}}, \xi_{a,\rm{o}}, \xi_{b,\rm{o}}$ represent the input pulse in frequency mode $a$, the output pulse in mode $a$, and the target output pulse in mode $b$. $\Lambda$ is the solved control pulse optimized for the mode conversion process.}
     \label{fig:mode_conv}
\end{figure} 

Specifically, we use an input pulse centered at 1550$\,$nm with a temporal width of $\tauG\equal 80\,$ps and $\kappa_{a,\rm{w}}\equal 4\OmG$, which ensures efficient absorption into cavity mode $a$. The output pulse has a target wavelength of 737~nm corresponding to the optical transition of the spin memory of choice (Section~\ref{subsec:spin_photon_interface}). Figure~\ref{fig:mode_conv} plots the input and output wave packets whose bandwidth is $\kappa_{b,\rm{w}}\equal 2\pi\times 200\,$MHz to match the bandwidth of the output pulse to the cavity containing the spin qubit. The small amplitude of the blue line in~Fig.~\ref{fig:mode_conv} indicates a very small fraction of the incident wave packet leaking through the cavity without being absorbed. This fraction decreases towards zero very rapidly as $\kappa_{a,\rm{w}}$ is increased relative to $\OmG$~\cite{Heuck_2020_PRA}.

\subsection{Spin-photon interface for quantum memory storage} \label{subsec:spin_photon_interface}

 Illustrated in Fig.~\ref{fig:qRX}, after passing through the MC, the photons at A and B each arrive at a bank of spin memories, which we take to be SiV$^-$ centers in diamond coupled to nanocavities as described in Ref.~\cite{Chen_2021} for memory multiplexing. Each SiV$^-$'s electron spin can map onto a neighboring nuclear spin for long memory storage via hyperfine interaction~\cite{Nguyen_2019_PRB} (Appendix~\ref{app_spin_cavity}), minimizing decoherence while the repeater waits for subsequent photons. For calculating the spin-spin Bell state fidelity, we assume a sufficiently long nuclear spin coherence time $\gg 1~$s and neglect dephasing errors.

 The spin memories can be optically interfaced through a $1\times N$ switching array that directly routes from 1 input channel to $N$ output channels. For an example, beam-steering devices such as a spatial light modulator may be used to route incoming photons to an arbitrary cavity in a free-space configuration. For the calculations covered in the remainder of the Article, we assumed using a sufficiently fast optical switch array with speed $\sim\sqrt{\eta}p_\text{ZALM}\sigma_\text{rep}\approx 100~$kHz. In Appendix~\ref{app_switch_array}, however, we also consider spin memories heterogeneously integrated into a LiNbO$_3$-based~\cite{Zhang_2017,Zhang_2021} PIC for high-bandwidth operations. For the latter, the entanglement generation rate decreases with increasing memory multiplexing due to a $\log_2 N $ depth Mach-Zehnder interferometer. One strategu to minimize the required dimension of the switching array is to assume the presence of multiple memories inside each cavity. Since the memories (at a given site) would be pre-characterized, their emission frequencies are known beforehand. The MC can then be used to spectrally select an emitter via upconversion.

    Importantly, prior to the switching array, the modes of polarization-encoded photonic state are split into two physical paths (on the PIC) by a polarization beam splitter. Only polarization mode $\ket{H}$ enters the $1\times N$ optical interposer to an array of $k$ nanocavities. The photon then reflects off the cavity and acquires a spin-state-dependent phase~\cite{Duan_2004}. Mode $\ket{V}$, on the other hand, acquires a constant phase by reflecting off a mirror in a path length-matched to that traveled by the $\ket{H}$ component. Together, the polarization-encoded qubit undergoes a controlled-phase gate that effectively entangles the photonic qubit with the spin memory. After the two polarization modes re-interfere at a beam splitter, they are subsequently detected in the diagonal basis to herald quantum teleportation, i.e. mapping a photonic qubit $\ket{\psi}_P=\alpha\ket{H}+\beta\ket{V}$ onto the spin qubit $\ket{\psi}_S=\alpha\ket{\downarrow}+\beta\ket{\uparrow}$.

Crucially, the controlled-phase gate fidelity depends on the atom-cavity cooperativity, $C$. In the limit of high cooperativity, the cavity reflection coefficient (for the $H$-polarization mode) would be
\begin{align}
    r \xrightarrow{C\gg 1} \frac{C-1}{C+1}
\end{align}
More details on the derivation can be found in Appendix~\ref{app_spin_cavity}. For the entanglement fidelity calculations presented in Section~\ref{sec:fidelity}, we assume a cooperativity of $C=100$~\cite{Bhaskar_2020}. Therefore, the mode-converted photon's bandwidth of 200~MHz is still much narrower than the spin's Purcell-broadened linewidth, a requirement which is paramount to the cavity reflection protocol~\cite{Duan_2004,Tiecke_2014}.

\section{Entanglement fidelity} 
\label{sec:fidelity}

We now analyze the spin-spin Bell state fidelity by accounting for imperfections in the qTX and qRX. First, for each spectral mode (i.e. spanning a single DWDM channel bandwidth), we compute the density matrix of the heralded photonic state $\rho_\mathrm{ZALM}$ based on Eq.~\ref{eqn:bigstatem}. $\rho_\mathrm{ZALM}$, in the polarization-Fock basis $\{\ket{H_A,V_A;H_B,V_B}\}$, contains up to two-photon contributions for $A$ and $B$. As for the spin qubits each initialized in a superposition state $\ket{\psi}_{S,i}=\ket{\downarrow}+\ket{\uparrow})/\sqrt{2}$, they form a product state
\begin{align}
    \rho_{S,i}=\left(\ket{\psi}\!\!\bra{\psi}_{S,i}\right)^{\otimes 2}
\end{align}
Hence, the collective initial state for both photonic and spin qubits is $\rho_i=\rho_\mathrm{ZALM}\otimes\rho_{S,i}$.

By considering finite cooperativity and waveguide-cavity coupling strengths based on the parameters used in Ref.~\cite{Chen_2021}, we evaluate the resultant controlled-phase gate operation acting only on the Bell basis states, $\hat{U_\mathrm{CZ}}$. For non-vacuum basis states outside of the Bell basis, we treat these contributions as erroneous photonic states that project the spins into maximally mixed states. Assuming perfect single-qubit gate for the Hadamard and Pauli-correction operations post-heralding, the teleported state's fidelity $\mathcal{F}_\mathrm{tele}$ shared between the two spin memories at A and B is calculated by taking the overlap with an ideal spin-spin Bell state $\ket{\Phi^+}$ (considering here only one of the four Bell states):
\begin{align}
    \mathcal{F}_\mathrm{tele} &= w_\mathrm{Bell}\cdot \langle\Phi^+|\widehat{U}_\mathrm{CZ}\rho_i \widehat{U}_\mathrm{CZ}^{\dagger}|\Phi^+\rangle + \frac{w_\mathrm{not-Bell}}{4}
\end{align}
where $w_\mathrm{Bell}$ and $w_\mathrm{not-Bell}$ represent the partial traces over the Bell and non-Bell basis states. The latter term is again due to terms outside of the Bell basis projecting the spin memories into maximally mixed states. Accounting for dark counts (later calculations assume $10^2~$Hz based on Ref.~\cite{Yin_2020}), we further modify the teleported state's fidelity to be,
\begin{align}
    \mathcal{F}_\mathrm{tele} &\rightarrow \left(w_\mathrm{Bell}\cdot \langle\Phi^+|\widehat{U}_\mathrm{CZ}\rho_i \widehat{U}_\mathrm{CZ}^{\dagger}|\Phi^+\rangle +\frac{w_\mathrm{not-Bell}}{4}\right)\nonumber\\
    &\quad \times (1-p_\mathrm{dark})^2+\frac{2(1-p_\mathrm{dark})p_\mathrm{dark}+p_\mathrm{dark}^2}{4}.
\end{align}

Next, we focus on infidelity caused by \textit{photon loss in the channel}. The issue arises from a competition between efficiency and fidelity. Since initializing spins is time-consuming~\cite{Bhaskar_2020}, time can be saved by not re-initializing the spins after every attempt (where photons transmit through atmosphere and arrived at qRX). A complication arises if a photon is lost \textit{after} reflecting off the cavity. This  qubit loss error projects the spin into a maximally mixed state; however since the loss is unheralded, the spins may not be re-initialized for the subsequent interacting photons. As a result, the \textit{average} spin-spin Bell state fidelity decreases as the number of attempt increases. By setting a maximum number of attempts $\Nmax$ before needing to re-initialize the spins, we can then minimize this photon loss infidelity, $\epsilon$. 

 To calculate the dependence of $\epsilon$ on $\Nmax$, we consider three scenarios when a photon arrives at the qRX. It can be (1) detected with probability $\sqrt{\eta}$ (assume identical downlink channels), (2) lost before reaching the spin with probability $\plost$, or (3) lost \textit{after} reaching the spin with probability $\pe$. Since generally $\plost\gg\{\sqrt{\eta},\pe\}$, we save time by not re-initializing the spins after every channel use.

However, as mentioned previously, skipping spin re-initialization after detector click potentially causes an unheralded error. Thus, we optimize the spin-spin entanglement fidelity by constraining the number of loading attempts before we re-initialize the spins. For simplicity of analysis, we fix an average qTX transmission rate at $1/\tau_0=p_\mathrm{ZALM}\sigma_\mathrm{rep}\approx 7.45$~GHz (Section~\ref{subsec:source_rate}). The probability of at least one unheralded error occurring in the first $m-1$ time bins conditioned on detector clicks on the $m\mathrm{th}$ bin is
\begin{align}
    P_{\mathrm{error}}(m, m)&=1-\left(\xi\plost/(1-\sqrt{\eta})\right)^{2(m-1)},
\end{align}
where $\xi$ is the probability of both A and B receiving photons, computed from taking a partial trace of $\rho_\mathrm{ZALM}$ over basis states with $\geq 1$ photons going to A and B. Then, the probability of error over $N$ bins is
\begin{align}
    {P}_\mathrm{error}&=\sum_{m=1}^N {P}_{\mathrm{error}}(m, m) (1-\eta)^{m-1}\eta.
\end{align}

Finally, the average state fidelity is
\begin{align}
    \mathcal{F} &= (1-P_\mathrm{error})\mathcal{F}_\mathrm{tele}+\frac{P_\mathrm{error}}{4}.
\end{align}
The maximum fidelity $\mathcal{F}_\mathrm{max}$ occurs when $P_\mathrm{error}=0$. We can then define $\epsilon=\mathcal{F}_{\mathrm{max}}-\mathcal{F}$ as the average infidelity arising from the unheralded photon loss error. Constraining $\epsilon$ therefore limits the maximum number of bins $\Nmax$ before needing to re-initialize the spin memories.

\section{Average entanglement generation rate} \label{sec:rate}

\begin{figure}
    \centering
    \includegraphics[width=0.5\textwidth]{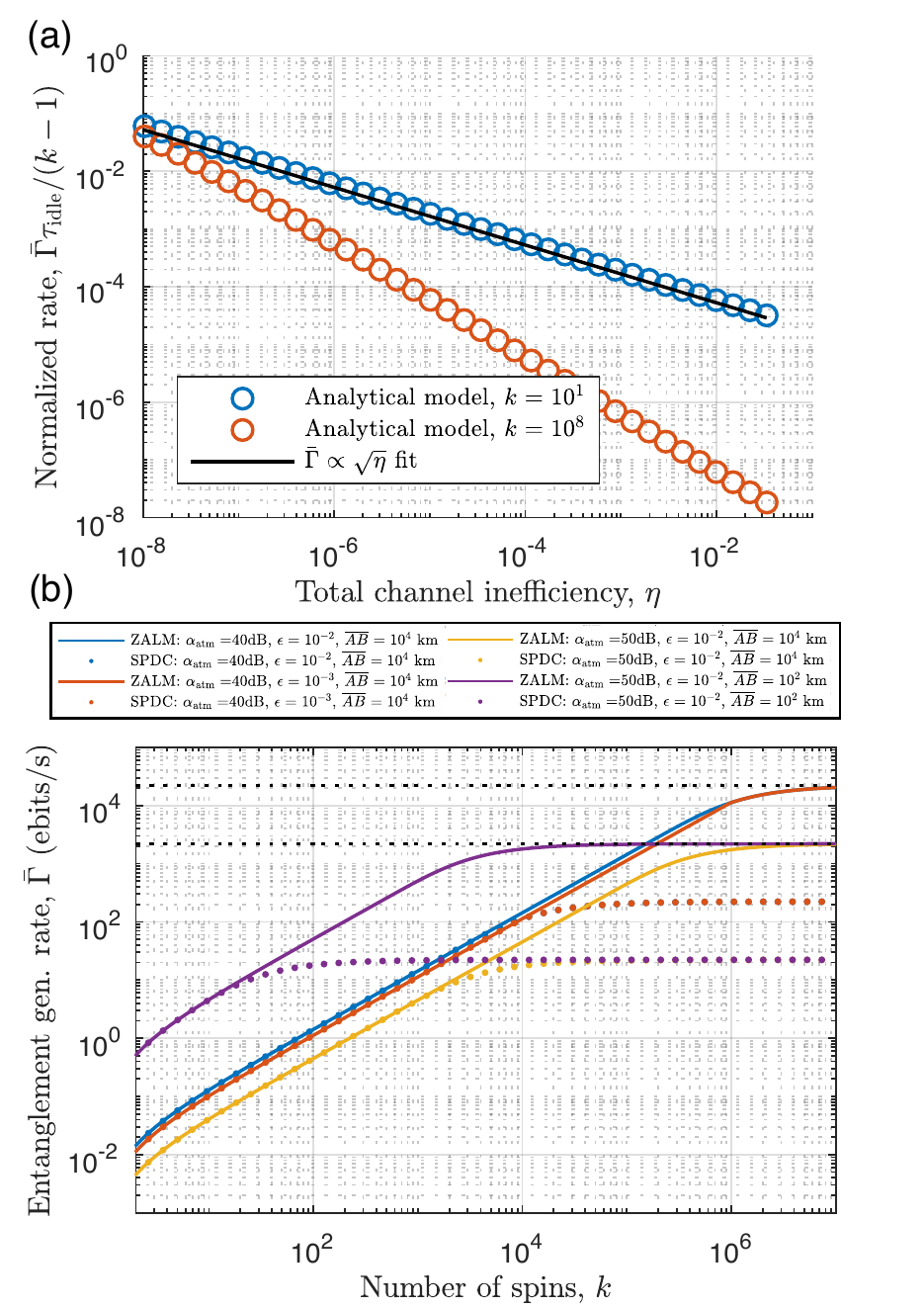}
    \caption{Entanglement generation rate $\bar{\Gamma}$ as a function of (a) the total channel efficiency $\eta$ and (b) the number of spins $k$. (a) In the memory-limited regime ($k=10^1$), the normalized rate $\bar{\Gamma}\tau_{\mathrm{idle}}/(k-1)$ scales $\propto\sqrt{\eta}$, whereas it scales $\propto\eta$ in the source-limited regime ($k=10^7$). (b) $\bar{\Gamma}$ vs $k$ for both ZALM and SPDC with varying ${\epsilon\in\{10^{-2},10^{-3}\}}$ and total downlink atmospheric attenuation ${\alpha_{\mathrm{atm}}=\{40,50\}~\mathrm{dB}}$. We additionally consider $\overline{AB}=10^2~$km (purple). These calculations assume a nuclear spin coherence time $\gg 1~$s.}
    \label{fig:k_rate}
\end{figure}

Lastly, we compute entanglement generation rate $\bar{\Gamma}$ by considering $k$ spins in the qRX. For simplicity, we let only a single initialized spin to accept photons at a time. Upon successful spin-photon mapping, A communicates with B to determine if B's corresponding photon was successfully detected. If both photons were detected, A then transfers the electron spin to the nuclear spin for memory storage (see Appendix~\ref{app_spin_cavity}). Otherwise, A re-initializes the spin qubit and awaits subsequent successful detection(s). This communicate-and-reset sequence takes a time $\tau_\mathrm{idle}$, given by the sum of communication time $\tau_\mathrm{comm}$ and the spin reset time $\tau_\mathrm{reset}$.

Since the remaining $(k-1)$ spins are inactive, each spin is ``on-duty'' for time $\tau_\mathrm{idle}/(k-1)$. The qTX generates $N_k\equiv\tau_\mathrm{idle}/[(k-1)\tau_0]$ attempts during a single spin's on-duty time. Furthermore, the target fidelity $\mathcal{F}=\mathcal{F}_{\mathrm{max}}-\epsilon(\Nmax)$ limits $\Nmax$, as mentioned in Section~\ref{sec:fidelity}. Hence, for a fixed $k$, the on-duty spin would actually be active for $N=\min(N_k,\Nmax)$ attempts.

In one clock cycle over time $\tau_\mathrm{idle}k/(k-1)$, each spin would re-initialize after every $N$ attempts. The probability both stations detect photons is then
\begin{align}
p_\mathrm{success} &= \eta  \left( \frac{1 - (1-\sqrt{\eta})^{2N}}{1 - (1-\sqrt{\eta})^2} \right),
\end{align}
yielding an average of $k p_\mathrm{success}$ successfully detected pairs. Finally, $\bar{\Gamma}$ is the ratio between the number of successfully detected pairs and clock cycle:
\begin{align}
\bar{\Gamma} &= \frac{p_{\mathrm{success}}\cdot(k-1)}{\tau_{\mathrm{idle}}}
\end{align}

Fig.~\ref{fig:k_rate}(a) compares the normalized rate $\bar{\Gamma}\tau_{\mathrm{idle}}/(k-1)$ at $k\in\{10^1,10^7\}$. In the memory-limited regime ($k=10^1$), $\bar{\Gamma}$ scales as $\sqrt{\eta}$ (see Appendix~\ref{app_sqrt_eta} for derivations), highlighting the advantage of the ``midpoint source''~\cite{Jones_2016} architecture. In contrast, in the source-limited regime ($k=10^8$), we recover the $\bar{\Gamma}\propto\eta$ scaling.

In Fig.~\ref{fig:k_rate}(b), we compare the entanglement generation rate $\bar{\Gamma}$ between using the ZALM BPS and using a free-running SPDC source (with a narrowband 200~MHz filter) as the qTX in both memory-limited and source-limited regimes. Furthremore, we compare their $\bar{\Gamma}$ between total downlink atmospheric attenuation $\alpha_{\mathrm{atm}}=40~$dB and 50~dB, corresponding to channel losses $\sqrt{\eta}\approx 0.2\%$ and $0.07\%$, respectively. The total loss accounts for attenuation in the downlink channel itself, 3.57~dB from adaptive optics ~\cite{Gruneisen_2017,footnote_adaptive_optics}, 3~dB from mode conversion inefficiency (Appendix~\ref{app freq conv}), 2.68~dB from the diamond nanocavity  (with cooperativity $C=100$), $\sim$0.8~dB insertion loss from the switching array (assumed a single-layer low-loss interposer), and 0.044~dB from detector inefficiency~\cite{Bhaskar_2020}. 

 We also consider infidelity arising from unheralded photon loss error $\epsilon=\{10^{-2},10^{-3}\}$, which effectively dictate the frequency of spin re-initialization (the consequential rate-fidelity trade-off is investigated in Appendix~\ref{app_rate_fidelity}). Lastly, we evaluate two ground-to-ground distances $\overline{AB}$ at $10^4$~km and $10^2$~km, with corresponding communication times $\tau_\mathrm{comm}=60$~ms and $\tau_\mathrm{comm}=0.5$~ms, respectively. The spin re-initialization time is assumed to be $\tau_{\mathrm{reset}}=30~\mu$s~\cite{Bhaskar_2020}.

For $\overline{AB}=10^4$~km (blue, orange, and yellow curves in Fig.~\ref{fig:k_rate}(b)), at low $k$, ZALM and SPDC have comparable $\bar{\Gamma}$ since the rate is limited by a quickly saturated bank of memories, i.e. photons are arriving at a rate faster than the spins are able to be freed up. However, as $k$ increases, the advantage of the ZALM BPS starts manifesting as its rate performance surpasses that of a non-heralded SPDC source.

\begin{figure}
    \centering
    \includegraphics[width=0.5\textwidth]{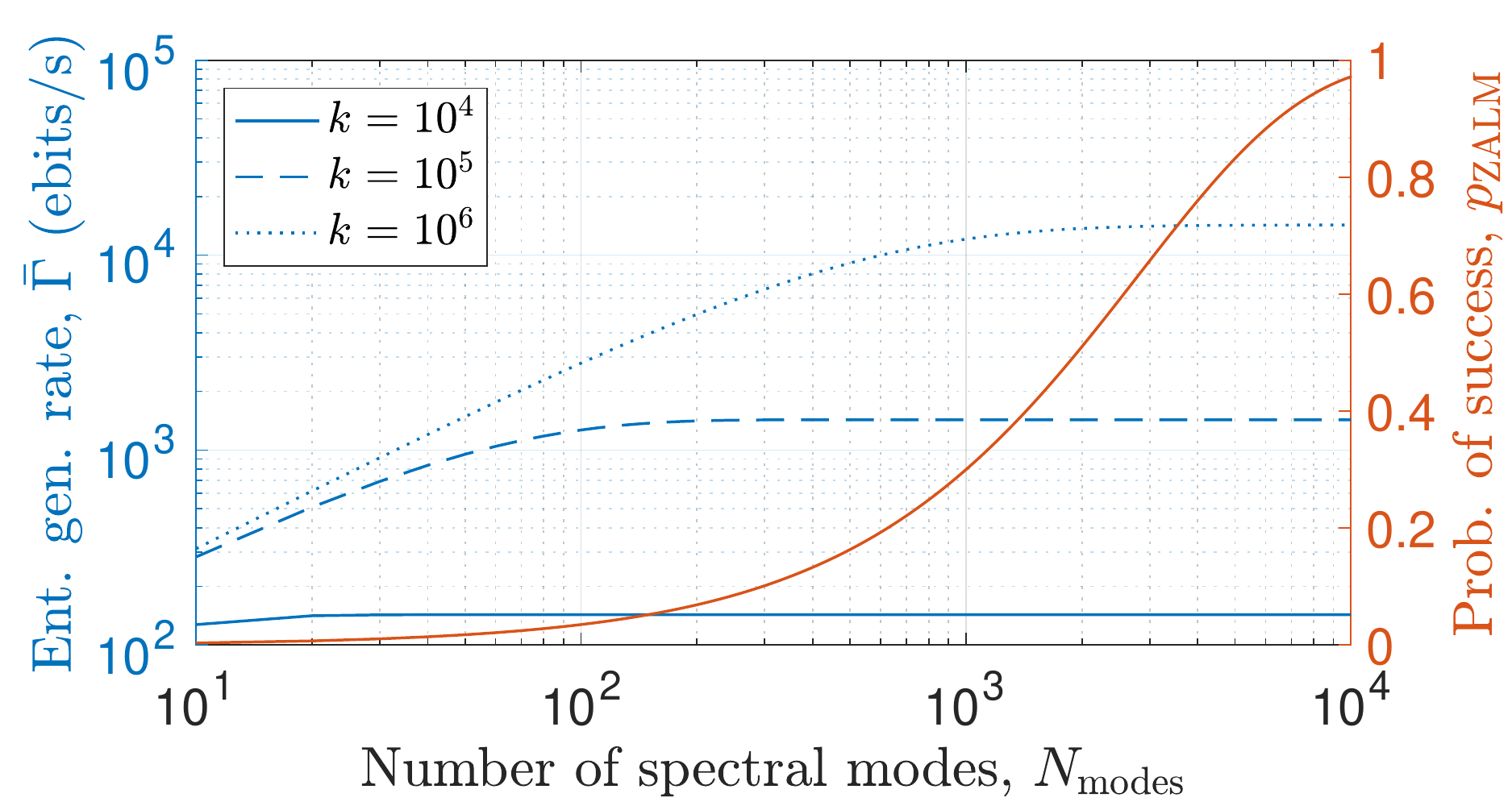}
    \caption{Entanglement generation rate $\bar{\Gamma}$ and the probability of success for the ZALM BPS $p_\text{ZALM}$ as a function of spectrally multiplexed modes in the qTX, $N_\text{modes}$. $\bar{\Gamma}$ is evaluated for $k=\{10^4,10^5,10^6\}$. Calculations assume $\eta=40~$dB, $\epsilon=10^{-2}$, and $\overline{AB}=10^4$~km.}
    \label{fig:rate_vs_numberModes}
\end{figure}

The point of divergence between the ZALM BPS and SPDC depends on $\epsilon$ and $\tau_{\mathrm{comm}}$. To maintain a small $\epsilon=10^{-3}$, A and B need to re-initialize the spins more frequently , i.e. reducing $\Nmax$ regardless of $k$. Hence, ZALM and SPDC achieve similar $\bar{\Gamma}$. Instead, if $\tau_{\mathrm{comm}}$ is reduced, e.g. smaller $\overline{AB}=10^2~$km, the effect of memory saturation is suppressed. As a result, ZALM would greatly outperform SPDC even with small $k$, shown by the purple curves in Fig.~\ref{fig:k_rate}(b). Lastly, $\bar{\Gamma}$ plateaus as $k$ increases indefinitely, at which point $\bar{\Gamma}$ is solely determined by $\eta$ and $\tau_0$,
\begin{align}
    \bar{\Gamma}\rightarrow \eta/(1-(1-\sqrt{\eta})^2)\cdot 2/\tau_0\cdot\log\left(1/(1-\sqrt{\eta})\right).
\end{align}
The upper bounds, indicated by the black dashed lines, are $7.3\times 10^2~$Hz for $\alpha_{\mathrm{atm}}=40~$dB and $7.3\times 10^1$~Hz $\alpha_{\mathrm{atm}}=50~$dB. Similar calculations for a ground-only quantum repeater network are presented in Appendix~\ref{app_ground_networks}, where we consider a midpoint source equidistant from A and B separated by $2L_\text{GG}=10^2$~km. Expectedly, a qTX based on the ZALM BPS shows a rate advantage by more than an order of magnitude than one based on a non-heralded SPDC source.

As shown in Fig.~\ref{fig:rate_vs_numberModes}, we further compute the entanglement generation rate $\bar{\Gamma}$ and the probability of success $p_\text{ZALM}$ as a function of the number of spectrally multiplexed modes $N_\text{modes}$ for $k=\{10^4,10^5,10^6\}$. The calculations here assume $\eta=40~$dB, $\epsilon=10^{-2}$, and $\overline{AB}=10^4$~km. Expectedly, $p_\text{ZALM}$ approaches unity as $N_\text{modes}$ increases. However, despite the increase in the qTX's transmission rate $1/\tau_0$, $\bar{\Gamma}$ quickly saturates at $N_\text{modes}<10^2$ for $k=10^4$, a manifestation of the memory-saturation effect. As $k$ increases to $10^6$, the rate at which spin memories is freed up at the qRX becomes comparable to $1/\tau_0$, giving rise to an increase in $\bar{\Gamma}$ monotonic with $N_\text{modes}$. Given the saturation behavior, the number of spin memories available at the qRX effectively sets the upper bound on $N_\text{modes}$ used in ZALM, past which point the rate advantage of the quasi-deterministic BPS nullifies. 

\section{Discussion and outlook} 
\label{sec:conclusion}

In this Article, we proposed a midpoint source architecture relying on a quasi-deterministic `zero-added-loss multiplexed' BPS for quantum networks. It leverages the large phase-matching bandwidth intrinsic to SPDC sources and utilizes spectral multiplexing via commercially available DWDMs. Moreover, our proposed mode converter crucially upconverts the heralded photons to spectral modes with matching frequency and bandwidth to those of solid-state spin qubits, a step which is essential for memory-based repeater networks. Our calculations show that the ZALM BPS greatly increases the entanglement generation rate $\bar{\Gamma}$, especially in low-transmission, memory-limited links. In our scheme,  upon receiving heralded photonic Bell pairs from the quantum transmitter on the satellite, the ground-based quantum receivers perform photon-to-spin mapping and herald spin-spin entanglement between two remote terrestrial stations. Our calculations show that in the memory-limited regime, our ``midpoint source'' scheme has a favorable $\sqrt{\eta}$ rate scaling. As memory multiplexing increases to $k\gtrapprox 10^2$, for ground-to-ground distance $10^2$~km, the ZALM BPS enables $\bar{\Gamma}>10$~Hz with theoretical spin-spin Bell state fidelity $\mathcal{F}>0.99$. This significantly outperforms the case where a single SPDC (non-heralded) source is used, in which $\bar{\Gamma}$ is lower by $\sim 2$ orders of magnitude. We stress that such a high bandwidth BPS should prove valuable to any two-way quantum repeater network configurations. For example, the advantage is apparent in a ground-only quantum network as shown in Appendix~\ref{app_ground_networks}. Additionally, the ZALM BPS may also benefit applications that must rely on distribution of optical entanglement, such as optical quantum computing~\cite{Kok_2007,Rudolph_2017,Zhong_2020}, precision measurement~\cite{Dowling_2009}, and all-optical quantum repeaters~\cite{Azuma_2015,Pant_2017}.

Currently, the upper bound of $\bar{\Gamma}$ is limited by the low probability of detecting the same Bell pair, which is a function of both channel loss $\eta$ and the BPS emission rate $1/\tau_0$.  While $\alpha$ (e.g. atmospheric attenuation) is often a fixed parameter, $\eta$ can be further improved by additional engineering: reducing the DWDM channel bandwidth to increase the number of spectrally multiplexed modes, minimizing loss in optical transmission (for e.g. increasing adaptive optics transmission efficiency, using larger optical apertures etc.), minimizing insertion loss for the $1\times N$ switching array, and increasing the cavity reflection efficiency (dictated by both design and fabrication capability currently~\cite{Knall_2022}). As for boosting the ZALM BPS' emission rate, using higher pump powers does increase the mean photon numbers $N_s$ (and consequently the heralding probability) at the cost of degraded state fidelity.  Having higher single photon detectioon efficiency on the satellite could also improve $1/\tau_0$. We stress that for ZALM's advantage to persist, the detection efficiency must be high in qTX entanglement swap and heralding. Missed photon detection translates to infidelity in the heralded state, countered only by lowering $N_s$, which further limits the heralding and entanglement distribution rate.

The implementation of the qTX in the proposal thus far demands having a large number of detectors in the midpoint source. Alternative to DWDM and having detectors for each spectral channel, one may also use time-of-flight measurements via dispersive optical elements already experimentally demonstrated in Ref.~\cite{Merkouche_2022_PRL}. With a detector reset time $\tau_r$ and a pump whose repetition rate of $\ll1/\tau_r$, a single detector would suffice, therefore significantly reducing the detector requirement for the BPS. However, we note that having a high detection efficiency is critical to correctly heralding Bell state creation in the qTX~\cite{Dhara_2021}. However, we note that there exists a large attenuation loss in the dispersive element ($\sim 10$~dB based on Ref.~\cite{Merkouche_2022_PRL}). Further technological improvements in constructing high-efficiency and highly dispersive optics are warranted to realize a practical qTX based on time-of-flight measurements.

 In the qRX, the mode converter is imperative to enabling $\ll 1$~ns long photons to interact with spin memories with $<$ns temporal width at frequencies 100s THz away. As opposed to sum-frequency generation in high-$Q$ ring resonators, time-lensing effect already realized in electro-optical modulating platforms (e.g. LiNbO$_3$) could also be viable alternatives~\cite{Karpinski_2017,Yu_2021}.

Given current demonstration of PIC-integrated $k\sim 10^2$ SiV$^{-}$~\cite{Wan_2020}, the achievable $\bar{\Gamma}$ is $\sim 1~$ebit/s. It remains another engineering challenge to  realize large-scale integration of solid-state spin qubits in photonic structures. Potential avenues may include implanting multiple emitters within each cavity via focused-ion beam~\cite{Schroder_2017,Wan_2020} and spectrally select a frequency-unique spin.

Lastly, instead of using a cavity-reflection spin-photon mapping protocol, BSM based on atomic emission interfering with photonics Bell pairs can also entangle remote spins~\cite{Jones_2016}.  The spin memories suitable for the proposed architecture extends to other matter qubits such as defects in Si~\cite{Redjem_2020} and SiC~\cite{Crook_2020}, rare-earth ions~\cite{Raha_2020,Askarani_2021}, atomic vapors~\cite{Trotta_2016}, cold atoms, and trapped ions, may be suitable memory qubits as well. Irrespective of the entanglement scheme and platform choice, our proposal remains viable as an optical entanglement distribution protocol  for midpoint source based quantum links.\newline

\section*{Acknowledgements}
We thank Zheshen Zhang (Univ. of Michigan), Paul G. Kwiat (UIUC), Eric Bersin (MIT-LL), Franco N. C. Wong (MIT), Babak N. Saif (GSFC, NASA), and Madison Sutula (MIT/Harvard) for discussions. We also would like to thank Thaddeus Ladd (HRL), Danny Kim (HRL), Sofiane Merkouche (HRL), Shanying Cui (HRL), and Shuoqin Wang (HRL) for invaluable feedback. K.C.C. acknowledges funding support by the National Science Foundation RAISE-TAQS (Grant No.1839155) and the MITRE Corporation Moonshot program. D.E. acknowledges partial support from the NSF RAISE TAQS program (Grant No.\ 1839155). P.D. acknowledges funding support from the Nicolaas Bloembergen Graduate Student Scholarship in Optical Sciences. M. H. acknowledges support from the Villum Foundation (QNET-NODES Grant No.\ 37417). W.D. is supported by the National Science Foundation to the Computing Research Association for the CIFellows 2020 Program. P.D., S.G., and D.E. acknowledge the National Science Foundation (NSF) Engineering Research Center for Quantum Networks (CQN), awarded under cooperative agreement number 1941583, for supporting this research. D.E. also acknowledges funding from the NSF C-ACCEL program.

During the preparation of our manuscript, we became aware of a similar work on heralded frequency-multiplexed BPSs~\cite{Merkouche_2022_PRL}.\newline

\section*{Conflicts of Interest}
S.G. has outside interests in Xanadu Quantum Technologies, Quantum Network Technologies, and SensorQ Technologies. These interests have been disclosed to the University of Arizona and reviewed in accordance with its conflict of interest policies. Any resulting conflicts of interest from these interests will be managed by The University of Arizona in accordance with its policies. D.E. holds shares in Quantum Network Technologies; these interests have been disclosed to MIT and reviewed in accordance with its conflict of interest policies, with any conflicts of interest to be managed accordingly.\newline

\clearpage
\onecolumngrid
\appendix
\renewcommand{\thefigure}{S\arabic{figure}}
\renewcommand{\theequation}{\thesection\arabic{equation}}
\setcounter{figure}{0} 

\renewcommand\thesection{\Alph{section}}
\renewcommand\thesubsection{\arabic{subsection}}


\section{Analysis of the Down Conversion Process}
\label{app_down_conv}
We consider a spontaneous parametric down-conversion (SPDC) Hamiltonian~\cite{Grice1997-rv,Grice2001-fk} in terms of the interacting field operators as
\begin{align}
    \hat{H}_{\text{int}}(t)=\int_{V} d^{3} r \chi^{(2)} \hat{E}_{P}^{(+)}(\mathbf{r}, t) \hat{E}_{S}^{(-)}(\mathbf{r}, t) \hat{E}_{I}^{(-)}(\mathbf{r}, t)+\text { H.c. }. \label{eqn:field}
\end{align}
The consituent terms in this expression are as follows 
\begin{itemize}
    \item[--] $\hat{E}_{j}(\mathbf{r}, t)$ $=\hat{E}_{j}^{(+)}(\mathbf{r}, t)+\hat{E}_{j}^{(-)}(\mathbf{r}, t)$ are the three interacting fields, with the mode label index $j=P, S, I$ identifying the pump, signal and idler fields respectively.
    \item[--] The crystal's nonlinearity is characterized by the second order nonlinear coefficient $\chi^{(2)}$. This coefficient is assumed to be equal over the frequency range of interest.
    \item[--] $V$ is the volume of interacting field regions in the nonlinear crystal. 
\end{itemize}
To simplify the analysis, the down-converted beams are constrained to be co-linear with the pump beam. The volume integral in Eq.~(\ref{eqn:field})  then becomes an integral over only one direction, which we choose to be $z$. The positive frequency part of the field operator $\hat{E}_{j}(z, t)$, is described by
\begin{align}
    \hat{E}_{j}^{(+)}(z, t)=\int d \omega_{j} A\left(\omega_{j}\right) \hat{a}_{j}\left(\omega_{j}\right) e^{i\left[k_{j}\left(\omega_{j}\right) z-\omega_{j} t\right]},
\end{align}
where $\hat{a}_{j}\left(\omega_{j}\right)$ is the photon annihilation operator for the mode defined by frequency $\omega_{j}$, the $z$ direction, and the polarization associated with the index j. The term $A\left(\omega_{j}\right)$  is a slowly varying function of frequency,
\begin{align}
A\left(\omega_{j}\right) =i \sqrt{\frac{\hbar \omega_{j} }{2 \epsilon_{0} n^{2}\left(\omega_{j}\right)\times V_j}}.
\label{eqn:svterm}
\end{align}
This term varies slowly w.r.t.~the field frequencies, and hence may be taken outside the integral. In Eq.~\eqref{eqn:svterm}, $ n(\omega_j) $ is the frequency dependent refractive index and $V_j$ is the mode volume. In case of a free-space implementation, this is calculated as the product of the crystal face area and the propagation length of the photons generated by the down-conversion process; the crystal face area is replaced by the waveguide mode area are in case of an on-chip implementation. 
 Since SPDC is a very inefficient process, the pump field must be relatively large. Accordingly, the electric-field operator $\hat{E}_{p}^{(+)}(\mathbf{r}, t)$ may be replaced by the classical field $E_{p}(\mathbf{r}, t)=\widetilde{\alpha}(t) e^{i k_{P}\left(\omega_{P}\right) z} $.  The interaction Hamiltonian may now be expressed as
\begin{align}
    \hat{H}_{\text{int}}(t)=& A \int_{0}^{L} d z \int d \omega_{I} \int d \omega_{S} \, \hat{a}_{I}^{\dagger}\left(\omega_{I}\right) \hat{a}_{S}^{\dagger}\left(\omega_{S}\right) \widetilde{\alpha}(t) \times e^{-i\left\{\left[k_{I}\left(\omega_{I}\right)+k_{S}\left(\omega_{S}\right)-k_{P}\left(\omega_{P}\right)\right] z-\left[\omega_{I}+\omega_{S}\right] t\right\}}+\mathrm{H} . \mathrm{c} . ,\label{eqn:interaction1}
\end{align}
where $L$ is the length of the crystal and $A\left(\omega_{j}\right)$ has been grouped into a single parameter $A$, along with several constants defined by 
\begin{align}
    A=A(\omega_S) A(\omega_I) \chi^{(2)} \approx A(\omega_P /2)^2 \chi^{(2)}
\end{align}

For a continuous wave (CW) pump, the nonlinear interaction prescribed by Eq.~(\ref{eqn:interaction1}) is a continuous process. We can argue that the interaction has therefore started long before the emission of signal and idler photons that we expect to arise as result. Thus we can extend the interaction time $ t_0\rightarrow-\infty $ and $t\rightarrow\infty $. 

With the revised limits of integration, the integral is somewhat easier to handle if the pump field is represented as its frequency components. This is in general true for any pump spectral shape~\cite{Rarity1994-tx}. We shall examine two cases in our analysis, (1) a CW pump of finite line width and (2) a pulsed mode-locked pump whose spectral characteristics are given as 
 \begin{enumerate}
 \item Continuous wave (CW) pump: The pump field for a CW laser may be expressed as $\tilde{\alpha}(t)=\int d\omega_P \, \alpha(\omega_P) \, e^{-i\omega_P t}$
 with spectral mode shape function $\alpha(\omega_P)$. We consider a  Gaussian spectral profile model for $\alpha(\omega_P) $ of the form,
 \begin{align}
     \alpha(\omega_P)=\exp(-(\omega_{P}-\omega_{P0})^2/2\sigma_P^2)
     \label{eq:cw_pumpshape}
 \end{align}
 where $\omega_{P,0}$ is the pump center frequency and $\sigma_P$ is the pump linewidth. 
 
     \item Mode-locked pump laser: For the present analysis the mode locked laser is modelled as a sum of CW-like laser lines separated by preset frequency separations given by 
    \begin{align}
    \begin{split}
        \alpha(\omega_P)
    &=\sum_{n=-N}^{N} \mathcal{S}(\omega_P)  \exp\left(\frac{-(\omega_{P}-\omega_{P,0}-n\Delta\omega_P)^2}{2\sigma_P^2}\right),
    \label{eq:ml_pumpshape}
    \end{split}
    \end{align}
    where $\mathcal{S}(\omega_P)$ is the overall mode envelope function for the laser. For a comb source with a Gaussian gain envelope, we choose  
$\mathcal{S}(\omega)=exp(-(\omega-\omega_{P0})^2/2 \sigma_{P,BW}^2)$ to be the standard functional form, where $\sigma_{P,BW}$ is the comb bandwidth. 
 \end{enumerate}
 
Irrespective of the pump shape, the integral in Eq.~(\ref{eqn:interaction1}) then becomes
\begin{align}
    \int_{t_{0}}^{t} d t^{\prime} \hat{H}_{\text{int}}\left(t^{\prime}\right)=& A \int_{-\infty}^{\infty} d t^{\prime} \int_{0}^{L } d z \int d \omega_{I}   \int d \omega_{S} \, \hat{a}_{I}^{\dagger}\left(\omega_{I}\right) \hat{a}_{S}^{\dagger}\left(\omega_{S}\right) \int d \omega_{P} \, \alpha\left(\omega_{P}\right) \nonumber\\
    & \times e^{-i\left\{\left[k_{I}\left(\omega_{I}\right)+k_{S}\left(\omega_{S}\right)-k_{P}\left(\omega_{P}\right)\right] z-\left[\omega_{I}+\omega_{S}-\omega_{P}\right] t\right\}} +\text { H.c. }
\end{align}
The time integral is performed first, yielding a $2 \pi \delta\left(\omega_{I}+\omega_{S}-\omega_{P}\right)$ term. Subsequently
integration over the length of the crystal yields,
\begin{align}
    \int_{t_{0}}^{t} d t^{\prime} \hat{H}_{\text{int}}\left(t^{\prime}\right)=& 2 \pi A \int d \omega_{I} \int d \omega_{S}  \, \hat{a}_{I}^{\dagger}\left(\omega_{I}\right) \hat{a}_{S}^{\dagger}\left(\omega_{S}\right) \times \alpha\left(\omega_{S}+\omega_{I}\right) \Phi\left(\omega_{S}, \omega_{I}\right)+\text { H.c. } \label{eqn:intH_final}
\end{align}
where $ \Phi(\omega_S,\omega_I) $  is the phase matching function (PMF) given by,
\begin{align}
    \Phi(\omega_S,\omega_{I})= \frac{\sin \left\{\left[k_{S}(\omega_{S})+k_{I}(\omega_{I})-k_{P}(\omega_{S}+\omega_{I})\right]L\right\}}{\left[k_{S}(\omega_{S})+k_{I}(\omega_{I})-k_{P}(\omega_{S}+\omega_{I})\right]L}
\end{align}

Using the concepts developed in the study of time-dependent perturbation theory, we may now work out a full Dyson series based expansion of the interaction to obtain the final state. For a generic interaction Hamiltonian $ H_{\text{int}}(t)$ acting on the initial state $\ket{\psi(t_0)}$ from $ t_0 $ to $ t $, the final state (upto a second order Dyson series expansion) can be expressed as,
\begin{align}
    \ket{\psi(t)}=\left[
    1+\frac{1}{i \hbar} \int_{t_0}^{t} dt' \, \hat{H}_{\text{int}} (t') +\frac{1}{(i\hbar)^2}\int_{t_0}^{t}  dt' \int_{t_0}^{t'} dt'' \hat{H}_{\text{int}}(t')  \hat{H}_{\text{int}}(t'')
    \right] \ket{\psi(t_0)}.
\end{align}
Such an expansion yields the following characteristic quantum state after the necessary integrals are evaluated.

\noindent \textit{Zeroth Order Term} --- The zeroth order approximation describes the regime without any nonlinear field interaction. Hence, this simply yields the broadband vacuum term $ \ket{0,0} $.
	
\noindent \textit{First Order Term }--- The first order Dyson expansion yields the term in which a single pair of signal and idler photons are created. Their spectral characteristics are governed by the joint spectral amplitude (JSA), $ J(\omega_S,\omega_I) =\alpha(\omega_S+\omega_I)\Phi(\omega_S,\omega_I)$ i.e. a product of the pump spectral function and the PMF.  The state may then be expanded as 
	\begin{subequations}
	    \begin{align}
		\ket{\psi^{(1)}}&=\frac{2\pi A}{i\hbar}  \int d \omega_{I} \int d \omega_{S}\, \alpha(\omega_{S}+\omega_{I}) \,  \Phi(\omega_{S},\omega_{I})\, \hat{a}_S^{\dagger} (\omega_S)\ \hat{a}_I^{\dagger} (\omega_I) \ket{0,0}\\
		&\equiv g^{(1)}\int\int  d \omega_{I}\, d \omega_{S}\; J(\omega_S,\omega_I)\, \hat{a}_S^{\dagger} (\omega_S)\ \hat{a}_I^{\dagger} (\omega_I)\ket{0,0}.
	\end{align}
	\end{subequations}
	Here, we make the simplification $g^{(1)}=2\pi A/i\hbar$.
It is important to note here that the JSA for the mode locked pump is a summation of JSA terms for the CW pump. Further, the mode spacing ($\Delta\omega_P$) is typically much smaller than the overall envelope's bandwidth, $\sigma_{P,BW}$; thus the overall JSA may be modelled by that of the pump envelope, rather the individual pump spectral lines.

\noindent \textit{Second Order Term }--- The second order term must account for time-ordering effects of subsequent down-conversion steps. Ref.~\cite{Branczyk2011-mg} has shown, that due to destructive interference, nontrivial terms in this expansion are eliminated. This leads us to calculate the second order terms as 
\begin{subequations}
    	\begin{align}
		\ket{\psi^{(2)}}&= \frac{1}{2}  \left(\frac{2\pi A}{i\hbar}\right)^2 \left( g'_0\ket{0,0}+\!\!
		\int d \omega_{I} \int d \omega_{S}  \int d \omega'_{I} \int d \omega'_{S} \,  J(\omega_S,\omega_I) \hat{a}_S^{\dagger} (\omega_S)\ \hat{a}_I^{\dagger} (\omega_I) \ket{0,0} J(\omega'_S,\omega'_I) \hat{a}_S^{\dagger} (\omega'_S)\ \hat{a}_I^{\dagger} (\omega'_I) \ket{0,0} \right)\\
		&\equiv g^{(2)} \left( g'_0\ket{0,0}+\int_{\text{all }\omega } J(\omega_S,\omega_I) J(\omega'_S,\omega'_I) \hat{a}_S^{\dagger} (\omega_S)\ \hat{a}_I^{\dagger} (\omega_I) \ket{0,0}  \hat{a}_S^{\dagger} (\omega'_S)\ \hat{a}_I^{\dagger} (\omega'_I) \ket{0,0}  \right)
		\label{eqn:perturb2}
	\end{align}
\end{subequations}
	where,
\begin{subequations}
    \begin{align}
        &g'_0=\int d\omega_I \int d\omega_S \, |\alpha(\omega_S+\omega_I) \Phi(\omega_S,\omega_I)|^2\\
        &g^{(2)}= 2(\pi A/i\hbar)^2=\left(g^{(1)}\right)^2/2
    \end{align}
\end{subequations}
Hence upto the considered second order expansion, the emitted state is expressed as,
\begin{align}
    \ket{\Psi}_{\mathrm{SPDC}}&=(1+g^{(2)} g'_0) \ket{\textbf{0}}+g^{(1)} \int J(\omega_{S},\omega_I)\,\hat{a}_S^{\dagger} (\omega_S)\, \hat{a}_I^{\dagger} (\omega_I) \ket{\textbf{0}}\nonumber\\
    &\quad+ 
    g^{(2)} \int J(\omega_{S},\omega_I) \, \hat{a}_S^{\dagger} (\omega_S) \,\hat{a}_I^{\dagger} (\omega_I) \cdot \, J(\omega_{S}',\omega_I') \hat{a}_S^{\dagger} (\omega_S') \,\hat{a}_I^{\dagger} (\omega_I') \ket{\textbf{0}}
    \label{eqn:DC_state2}
\end{align}
where $\ket{\textbf{0}}$ represents the vacuum state. Note that this state is unnormalized and hence contains information about the total emission rate of the source.

\subsection{Post-Measurement State and Detection Statistics of Zero-Added Loss Multiplexed Source}
 The output state for a single entangled pair source (comprised of two down conversion processes) is
\begin{align}
    \ket{\Gamma}=&\ket{\Psi}_{\mathrm{SPDC}}^{\otimes 2}= \left(\ket{0,0}+\ket{\psi^{(1)}}+\ket{\psi^{(2)}}\right)^{\otimes 2}\\
    =&\ket{0,0}^{(1)}\ket{0,0}^{(2)}+\ket{\psi^{(1)}}^{(1)} \ket{0,0}^{(2)}+\ket{0,0}^{(1)}\ket{\psi^{(1)}}^{(2)} \text{(vacuum + Bell pair )} \nonumber\\
    &+\ket{\psi^{(2)}}^{(1)} \ket{0}^{(2)} + \ket{0}^{(1)}  \ket{\psi^{(2)}}^{(2)} + \ket{\psi^{(1)}} ^{(1)}\ket{\psi^{(1)}} ^{(2)}  \quad \text{(two photon terms + vacuum contri.)} \nonumber\\
    &+\ldots+\ket{\psi^{(2)}}^{(1)} \ket{\psi^{(2)}}^{(2)}  \quad \text{(higher order terms)}
\end{align}
We clarify on the mode labelling convention used in the generalized analysis of this Appendix. Each SPDC process generates a signal-idler pair; an entangled pair source is modelled to comprise of two SPDC processes. Let us label the signal and idler as $S$ and $I$ respectively; additionally let us associate the subscript (say $k$) to the entangled pair source. The superscript is used to label the signal-idler pair (or equivalently, which constituent SPDC process). So the mode label $S^{(2)}_\mfa$ signifies the signal of the second SPDC process for the entangled pair source labelled $\mathfrak{a}$.

\begin{figure}[ht!]
    \centering{\includegraphics[width=0.5\linewidth]{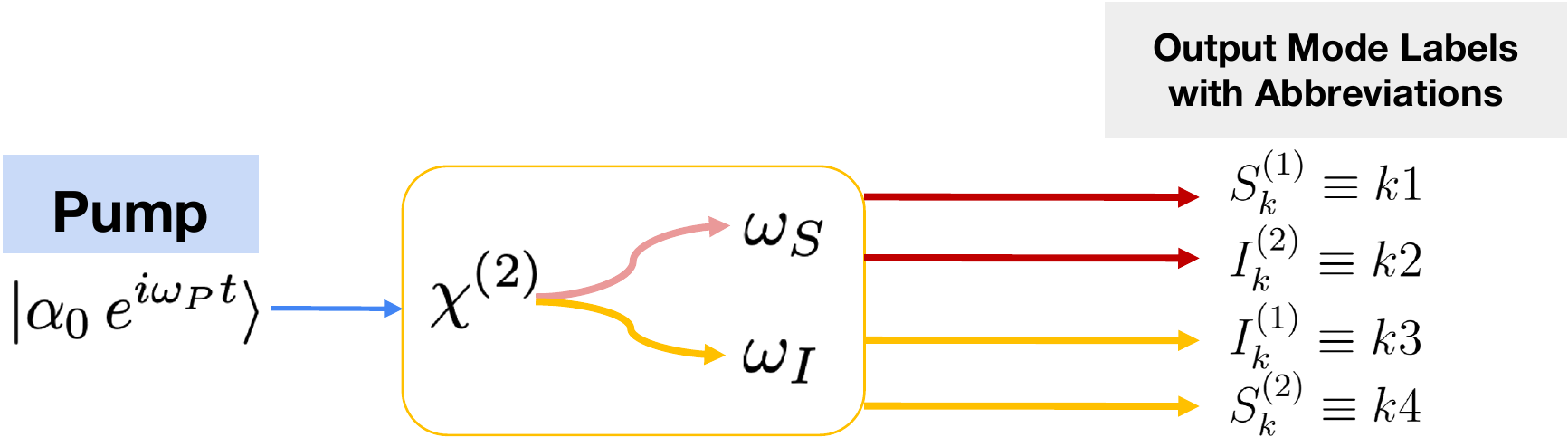}}
    \caption{Schematic of the individual SPDC entangled pair source using a CW pump excitation.}
    \label{fig:diag_spdc}
\end{figure}

Since the idler beams are swapped, we can assume the mode ordering to be $ S^{(1)}_k, I^{(2)}_k, S^{(2)}_k,I^{(1)}_k $ to get the standard state expansion, where $ k =\mfa,\mfb$ determines which entangled pair source (of two) we are referring to. Additionally, we may simplify the mode notation by adopting the abbreviations proposed in Fig.~\ref{fig:diag_spdc}.

\begin{figure}[ht!]
    \centering{\includegraphics[width=0.75\linewidth]{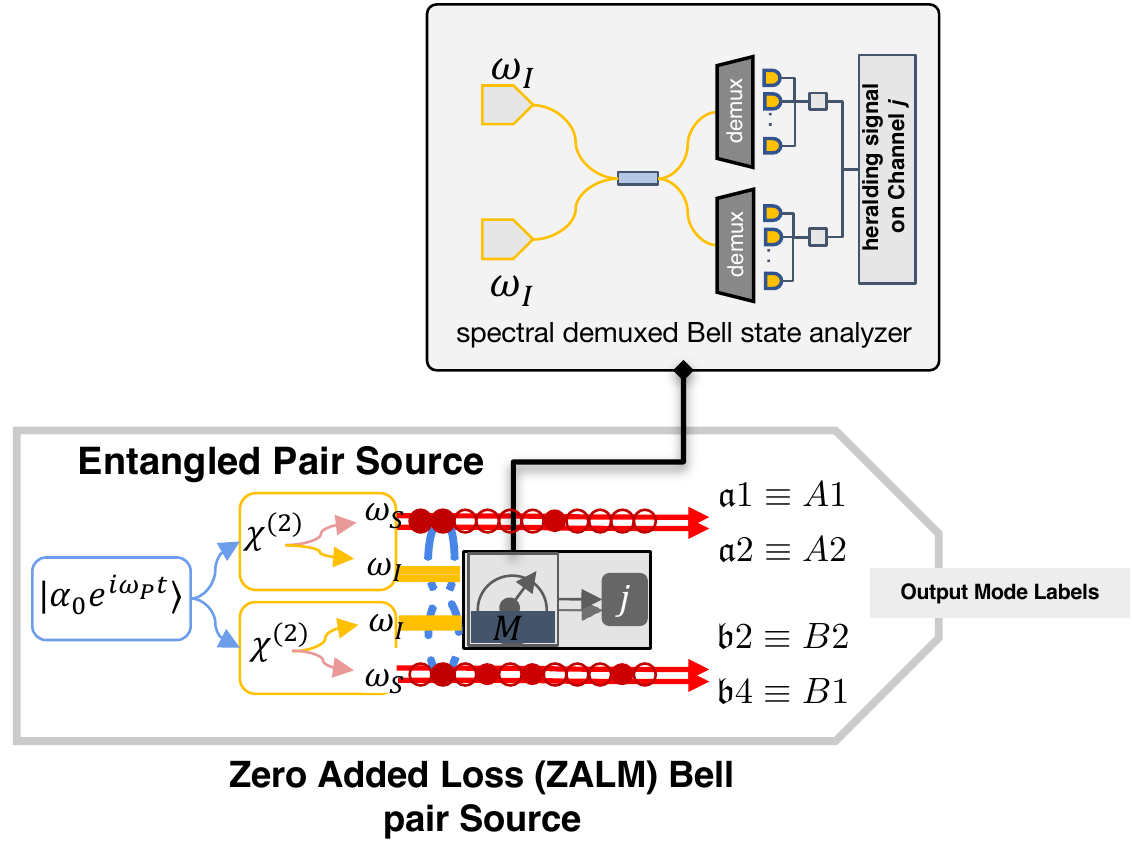}}
    \caption{Schematic of the Zero added loss Multiplexed (ZALM) Bell pair source.}
    \label{fig:diag_demuxBSM}
\end{figure}

The spectrally demultiplexed Bell state analyzer in  Fig.~\ref{fig:diag_demuxBSM} governs the mode interactions; we shall analyze a subset of them to determine the final quantum state. As per our abbreviated  notation, the modes labelled by ${\mfa4} $ and $\mfa3$ interact with modes $\mfb1$ and $\mfb2$, respectively. Under the action of the balanced beamsplitters preceeding the detection, the creation operators are transformed as per the following rule,
\begin{subequations}
    \begin{align}
        \hat{a}^\dagger_{D_1}:=\frac{1}{\sqrt{2}} \left( \hat{a}^\dagger_{\mfa4} + \hat{a}^\dagger_{\mfb 1}\right),
        \hat{a}^\dagger_{D_2}:=\frac{1}{\sqrt{2}} \left( \hat{a}^\dagger_{\mfb 1} - \hat{a}^\dagger_{\mfa4}\right),\\
        \hat{a}^\dagger_{D_3}:=\frac{1}{\sqrt{2}} \left( \hat{a}^\dagger_{\mfa 3} + \hat{a}^\dagger_{\mfb 2}\right),
        \hat{a}^\dagger_{D_4}:=	\frac{1}{\sqrt{2}} \left( \hat{a}^\dagger_{\mfb 2} - \hat{a}^\dagger_{\mfa 3}\right),
    \end{align}
\end{subequations}
where $\hat{a}^\dagger_{D_i}$ is the creation operator for corresponding detector bank. 	Subsequently, we may analyze the modes pairwise to determine the quantum state emitted in a complementary pair of undetected modes. As an example, consider the $\mfa4 \Leftrightarrow \mfb 1$ interaction; the corresponding complementary modes are $\mfa 2$ and $ \mfb 3$ respectively. Detection of a $1,0$ click pattern on the pair of modes i.e. detectors $D_1$ and $D_2$ on one of the uDWDM channels imposes a spectral window on the detected photon. The detection jitter (which is a detector parameter) in conjunction with the spectral window, govern the temporo-spectral shape of the heralded photon pair. Given a spectral filter of width $\Delta \Omega$~Hz and a detector with a detection jitter of $\delta\tau$~s, the temporal extent of the un-detected photons is given by $\max(\delta \tau,1/\Delta\omega)$~s.

We consider detection of photons in modes $\mfa4$ and $\mfb1$ overall a spectral range $\omega_{\mfa4},\omega_{\mfb1}\in (\Omega, \Omega') $. This heralds the quantum state $\ket{\varphi}$ on the undetected modes
\small{
    \begin{align}
\ket{\varphi} \propto
    \biggl(&\int d\omega_{\mfa 2} \int_{\Omega}^{\Omega'} d\omega_{\mfa 4}\,  \alpha\left(\omega_{\mfa 2}+\omega_{\mfa 4}\right)\Phi\left(\omega_{\mfa 4},\omega_{\mfa 2}\right) \hat{a}^\dagger_{\mfa 2} (\omega_{\mfa2}) +\int d\omega_{\mfb 3} \int_{\Omega}^{\Omega'} d\omega_{\mfb 1}\, \alpha\left(\omega_{\mfb 1}+\omega_{\mfb 3}\right)\Phi\left(\omega_{\mfb 1},\omega_{\mfb 3}\right) \hat{a}^\dagger_{\mfb 3} (\omega_{\mfb 3})\biggr) 
        \ket{0}_{\mfa 2}\ket{0}_{\mfb 3}.
        \label{eqn:state1}
\end{align}
}
We note that modes $\mfa2$ and $\mfb3$ span a complementary frequency range i.e.\ $\omega_{\mfa2},\omega_{\mfb3}\in(\omega_P-\Omega',\omega_P-\Omega)$ ensures energy conservation and phase-matching of the down-converted beams. Similarly, the detection of a $1,0$ pattern on the detector banks $ D_3 $ and $ D_4 $ for a the detection channel  $\omega_{\mfa3},\omega_{\mfb2}\in(\Omega,\Omega')$ heralds a quantum state $\ket{\varphi'}$ similar to Eq.~(\ref{eqn:state1}) above, and is described by
\small{\begin{align}
    \ket{\varphi'} \propto	&\biggl(\int d\omega_{\mfa 1} \int_{\Omega}^{\Omega'} d\omega_{\mfa 3}\,  \alpha\left(\omega_{\mfa 1}+\omega_{\mfa 3}\right)\Phi\left(\omega_{\mfa 3},\omega_{\mfa 1}\right) \hat{a}^\dagger_{\mfa 1}(\omega_{\mfa 1}) +\int d\omega_{\mfb 4} \int_{\Omega}^{\Omega'} d\omega_{\mfb 2}\, \alpha\left(\omega_{\mfb 2}+\omega_{\mfb 4}\right)\Phi\left(\omega_{\mfb 2},\omega_{\mfb 4}\right) \hat{a}^\dagger_{\mfb 4} (\omega_{\mfb 4})\biggr)
    \ket{0}_{\mfa 1}\ket{0}_{\mfb 4}.
    \label{eqn:state2}
\end{align}
}
Hence, the complete detection pattern of $ 1,0,1,0 $ on detectors $D_1$---$D_4$, yields the final state $ \ket{\varphi}\otimes\ket{\varphi'} $, which we may express compactly as $\ket{\varphi}\otimes\ket{\varphi'}\equiv	 \ket{\varPsi}  +\ket{\Xi}$
with the constituent terms,
\begin{subequations}
\begin{align}
    \ket{\varPsi}= \int d\bar{\omega} \left( J(\omega_{\mfa1},\omega_{\mfa 3})  \cdot J(\omega_{\mfb1},\omega_{\mfb 3})\, \hat{a}^\dagger_{\mfa 1}(\omega_{\mfa1})\,  \hat{a}^\dagger_{\mfb 3} (\omega_{\mfb3})
    + J(\omega_{\mfa2},\omega_{\mfa 4})  \cdot J(\omega_{\mfb2},\omega_{\mfb 4})\, \hat{a}^\dagger_{\mfa 2} (\omega_{\mfa 2})\, \hat{a}^\dagger_{\mfb 4} (\omega_{\mfb 4})
    \right)  \ket{0}_{\mfa 1}  \ket{0}_{\mfa 2}  \ket{0}_{\mfb 3} \ket{0}_{\mfb 4} \\
    \ket{\Xi}= \int d\bar{\omega} \left( J(\omega_{\mfa1},\omega_{\mfa 3})  \cdot J(\omega_{\mfa2},\omega_{\mfa 4})  \, \hat{a}^\dagger_{\mfa 1}(\omega_{\mfa 1})\, \hat{a}^\dagger_{\mfa 2} (\omega_{\mfa 2})
    +  J (\omega_{\mfb1},\omega_{\mfb 3}) \cdot J (\omega_{\mfb2},\omega_{\mfb 4})\, \hat{a}^\dagger_{\mfb 3} (\omega_{\mfb 3}) \,\hat{a}^\dagger_{\mfb 4} (\omega_{\mfb 4})
    \right) \ket{0}_{\mfa 1}   \ket{0}_{\mfa 2}  \ket{0}_{\mfb 3} \ket{0}_{\mfb 4}.
\end{align}
\end{subequations}
Here, $ J(\omega_i,\omega_j) =\alpha(\omega_i+\omega_j)\Phi(\omega_i,\omega_j) $, represents the joint spectral amplitude function for the signal-idler frequency pair $ \omega_i,\omega_j $. The $ \ket{\varPsi} $ component is the spectral equivalent of a broadband Bell pair of the form $(\ket{1,0}_A\ket{0,1}_B+\ket{0,1}_A\ket{1,0}_B)/\sqrt{2}$. In contrast, $ \ket{\Xi} $ is a state in which either Alice or Bob receive both photons; this is equivalent to a broadband $(\ket{1,1}_A\ket{0,0}_B+\ket{0,0}_A\ket{1,1}_B)/\sqrt{2}$ state. Note that the latter component is a spurious term which limits the fidelity of the generated entangled state. 
The overall general state may be expressed as 
\begin{align}
    \ket{S}\propto 	 \int d\bar{\omega} \Biggl[ & \biggl( J(\omega_{\mfa1},\omega_{\mfa 3})  \cdot J (\omega_{\mfb1},\omega_{\mfb 3})\, \hat{a}^\dagger_{\mfa 1} (\omega_{\mfa 1}) \, \hat{a}^\dagger_{\mfb 3} (\omega_{\mfa 3})
    +(-1)^{m_1} J(\omega_{\mfa2},\omega_{\mfa 4})  \cdot J (\omega_{\mfb2},\omega_{\mfb 4})\, \hat{a}^\dagger_{\mfa 2} (\omega_{\mfa 2})\, \hat{a}^\dagger_{\mfb 4} (\omega_{\mfb 4})\biggr)\nonumber \\ 
    &+(-1)^{m_2}\biggl( J(\omega_{\mfa1},\omega_{\mfa 3})  \cdot J(\omega_{\mfa2},\omega_{\mfa 4})  \, \hat{a}^\dagger_{\mfa 1}(\omega_{\mfa 1})\, \hat{a}^\dagger_{\mfa 2}(\omega_{\mfa 2})
    +  (-1)^{m_1} J (\omega_{\mfb1},\omega_{\mfb 3}) \cdot J (\omega_{\mfb2},\omega_{\mfb 4})\, \hat{a}^\dagger_{\mfb 3} (\omega_{\mfb 3})\, \hat{a}^\dagger_{\mfb 4} (\omega_{\mfb 4})\biggr)
    \Biggr] \nonumber\\
    &\times \ket{0}_{\mfa 1}  \ket{0}_{\mfa 2}  \ket{0}_{\mfb 3} \ket{0}_{\mfb 4}
    \label{eqn:bigstate}
\end{align}
where the compressed integral $ \int d\bar{\omega} $ denotes the limited frequency integrals (for the corresponding spectral channels)
\begin{align}
    \int d\bar{\omega}\equiv \int d\omega_{\mfa 1}\int d\omega_{\mfa 2}  \int_{\Omega}^{\Omega'} d\omega_{\mfa 3} \int_{\Omega}^{\Omega'} d\omega_{\mfa 4} \int_{\Omega}^{\Omega'} d\omega_{\mfb 1} \int_{\Omega}^{\Omega'} d\omega_{\mfb 2} \int d\omega_{\mfb 3} \int d\omega_{\mfb 4}
\end{align}
The frequency extent of the undetected modes $(\mfa1,\mfa2,\mfb3,\mfb4)$ are restricted to a $(\omega_P-\Omega',\omega_P-\Omega)$ by the detection windows (neglcting the pump linewidth). In Eq.(\ref{eqn:bigstate}), $ m_1 $ and $ m_2 $ are parity bits given by Table~\ref{tab:parity}.
\begin{table}[]
    \begin{tabular}{cccccc}
        \toprule
        \multicolumn{4}{c}{Click Pattern} & \multicolumn{2}{c}{Parity Bits}\\ \cmidrule(r){1-4} \cmidrule(r){5-6}
        $D_1$  & $D_2$  & $D_3$  & $D_4$  &         $m_1$               &     $m_2$                   \\ \midrule
        1&    0    &   1    &     0   &     0       &      0         \\
        0&    1    &   0   &     1   &     0       &      1           \\
        0&    1    &   1    &     0   &     1       &      1           \\
        1&    0    &   0   &     0   &     1       &      0           \\ \bottomrule
    \end{tabular}
    \caption{Table of parity bits $(m_1,m_2)$ for a given click pattern in detectors $D_1-D_4$}
    \label{tab:parity}
\end{table}
 For the purposes of the current protocol, we translate to a qubit notation for the modes that make up the quantum state transmitted. We choose the equivalent naming convention where  $ \mfa1,\mfa 2,\mfa3,\mfa4 \equiv A1,A2,A1',A2'$ respectively and $\mfb1,\mfb2,\mfb3, \mfb4 \equiv B2',B1', B2,B1$. The ordering for the $\mfb$ modes are reversed to match the mode ordering in Fig.~\ref{fig:diag_demuxBSM} where $\mfb3 $ and $\mfb 4$ represent modes transmitted to Bob. With this mode relabelling, Eq.~(\ref{eqn:bigstate}) becomes, 
	\begin{align}
		\ket{S}\propto 	 \int &d\bar{\omega} \Biggl[  \biggl( J(\omega_{A1},\omega_{A1'})  \, J (\omega_{B2'},\omega_{B2})\, \hat{a}^\dagger_{A 1} (\omega_{A 1}) \, \hat{a}^\dagger_{B2} (\omega_{B2})
		+(-1)^{m_1} J(\omega_{A2},\omega_{A2'})  \, J (\omega_{B1'},\omega_{B1})\, \hat{a}^\dagger_{A 2} (\omega_{A2})\, \hat{a}^\dagger_{B1} (\omega_{B1})\biggr)\nonumber \\ 
		&+(-1)^{m_2}\biggl( J(\omega_{A1},\omega_{ A1'}) \,J(\omega_{A2},\omega_{A2'})  \, \hat{a}^\dagger_{A1}(\omega_{A1})\, \hat{a}^\dagger_{A2}(\omega_{A2})
		+  (-1)^{m_1} J (\omega_{B2'},\omega_{B2}) \,J (\omega_{B1'},\omega_{B1})\, \hat{a}^\dagger_{B1} (\omega_{B1})\, \hat{a}^\dagger_{B2} (\omega_{B2})\biggr)
		\Biggr] \nonumber\\
		&\times \ket{0}_{A1}  \ket{0}_{A2}  \ket{0}_{B1} \ket{0}_{B2}
		\label{eqn:bigstate2}
	\end{align}

We use the equivalent notation $\ket{S}\equiv\ket{S(\omega_{P0},\sigma_P)}$ to denote that the state specified in Eq.~(\ref{eqn:bigstate2}) arises from a CW pump field of the form Eq.~(\ref{eq:cw_pumpshape}). Hence given our choice of the mode-locked pump in Eq.~(\ref{eq:ml_pumpshape}), the multiplexed heralded state can be succinctly described by
\begin{align}
    \ket{\mathbf{S}}=\sum_{n =-N}^{N} \ket{S(\omega_{P0}+n\Delta\omega_P,\sigma_P)}.
\end{align}
This state poses additional changes; since there are multiple `center frequencies', the signal and idler photons have an added degree of uncertainity in their origin. Considering a single SPDC source (say modes labelled by $\mathfrak{a}$) a signal-idler pair (say at frequencies $\omega_{a1}$ and $\omega_{a3}$)  could be generated by each pump line (centered at $\omega_P=\omega_{P0}+n\Delta\omega_P$). The contribution of each pump line is determined by the spectral envelope ($\mathcal{S}(\omega_P)$) of the pump and more crucially, by the joint spectral amplitude $J(\omega_{\mathfrak{a}_1},\omega_{\mathfrak{a}_3})$ function.

\section{Heralded photonic Bell state} \label{app_photonic_bell_state}
\begin{figure*}[t]
    \centering
    \includegraphics[width=1\textwidth]{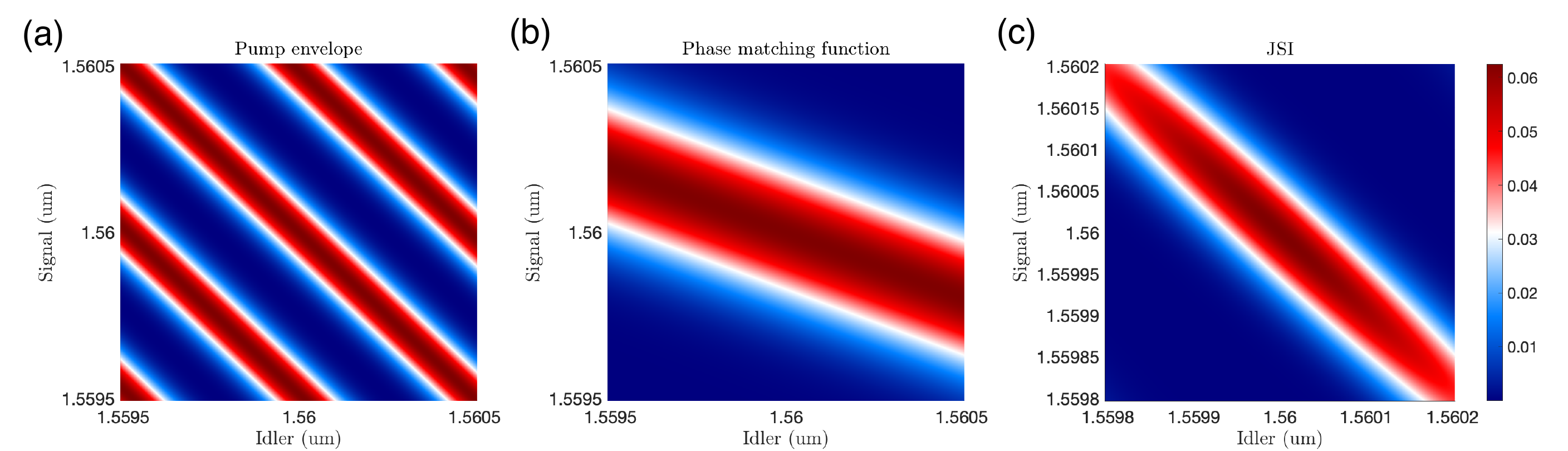}
    \caption{Numerically calculated (a) pump envelope function with a frequency comb source, (b) phase matching function using the Gaussian approximation, and the resultant (c) joint spectral intensity (JSI) function (modulus square of the JSA function). The plot axes for (a) and (b) spans across a $10\times 12.5~$GHz, with the center wavelength at $2\times 2\pi c/\omega_p=1560~$nm. For (c), the axes span $12.5~$GHz corresponding to the spectral width of each DWDM channel.}
    \label{fig:supp_JSI_beforeDWDM}
\end{figure*}

The heralded photonic Bell state fidelity may suffer from timing uncertainty in the detection and imperfect mode conversion. Here we address the aforementioned issues through an example of calculated joint spectral amplitude (JSA) function $J$, which is a product of the pump envelope $\alpha(\omega_P)$ and the phase matching function $\Phi(\omega_S,\omega_I)$ (Appendix~\ref{app_down_conv}). We first assume a repetition rate of {1~GHz} for the mode-locked pump laser, with each comb width being 12.5~GHz (matching that of the DWDM channel as explained later). We also take the SPDC material to be magnesium-oxide doped LiNbO$_3$. Using the Sellmeier equations for both the signal (extraordinary) and the idler (ordinary) photons, and applying a Gaussian approximation~\cite{Grice2001-fk}, we can compute the phase matching function $\Phi(\omega_S,\omega_I)$. Multiplying $\alpha(\omega_P)$ with $\Phi(\omega_S,\omega_I)$, we then obtain the JSA function. Figure~\ref{fig:supp_JSI_beforeDWDM} shows the three functions respectively.

\begin{figure*}[t]
    \centering
    \includegraphics[width=0.7\textwidth]{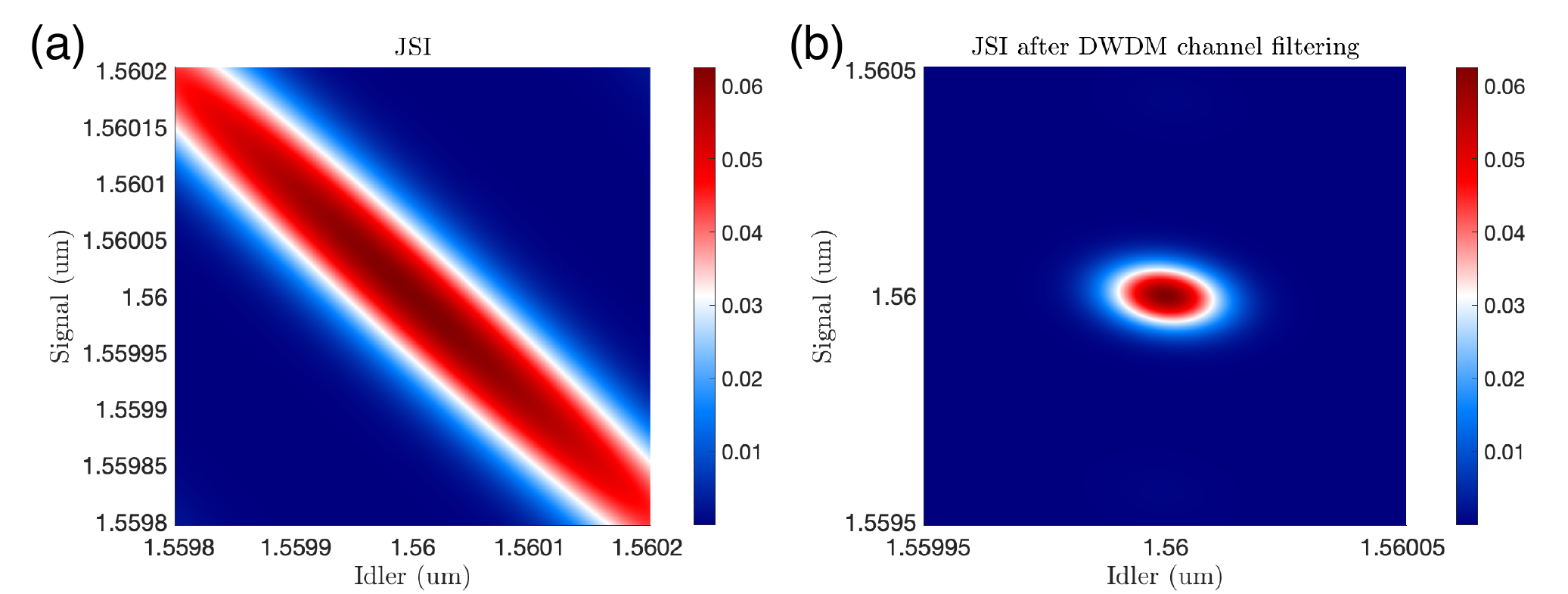}
    \caption{The JSI function before and after applying spectral filtering via a Gaussian-shaped DWDM channel profile. The calculated entropy based on Schmidt decomposition is $S=0.446$.}
    \label{fig:supp_JSI_afterDWDM}
\end{figure*}

We further assume a Gaussian spectral profile for the DWDM channel that shapes the JSA, as illustrated in Fig.~\ref{fig:supp_JSI_afterDWDM}. To quantify the level of separability, we perform Schmidt decomposition on the JSA to extract its eigenvalues $\lambda_m$ with orthonormal basis vectors $\{u_m(\omega)\},\{v_m(\omega)\}$~\cite{Grice2001-fk}:
\begin{align}
    J(\omega_s,\omega_i) &= \sum_m \sqrt{\lambda_m} u_m(\omega_s) v_m(\omega_i)
\end{align}
where
\begin{align}
    \int K_1(\omega,\omega') u_m(\omega') d\omega' &= \lambda_m u_m(\omega)\\
    \int K_2(\omega,\omega') v_m(\omega') d\omega' &= \lambda_m v_m(\omega)\\
    K_1(\omega,\omega') &= \int J(\omega,\omega_2) (J(\omega',\omega_2))^* d\omega_2\\
    K_2(\omega,\omega') &= \int J(\omega_1,\omega) (J(\omega_1,\omega'))^* d\omega_1
\end{align}

The entropy of entanglement, which approaches zero as the JSA becomes more separable, is calculated based on:
\begin{align}
     S &= -\sum_{k=0}^\infty \lambda_k \log_2 \lambda_k
\end{align}
    
If the inverse of the detector jitter time is much greater than the DWDM channel bandwidth of 12.5~GHz, then the infidelity of the photonic Bell state stemming from projection into a mixed state is negligible. For calculations presented in the main text, we assumed a timing uncertainty of 1~ps~\cite{Korzh_2020} corresponding to 1~THz, which greatly exceeds 12.5~GHz. However, even with ideal detection, the heralded photonic Bell state composed by the signal photons from two SPDC sources may still have a complex spectral shape, which may lead to infidelity caused by the mode conversion step. This is largely a consequence of using a frequency comb source with a repetition rate comparable to the DWDM channel width.

Two avenues to circumvent this issue are (i) injecting a complex pump pulse shape in the MC corresponding to the spectral shape in the heralded Bell state or (ii) engineering the JSA function~\cite{Grice2001-fk}. For the calculated example shown in Fig.~\ref{fig:supp_JSI_beforeDWDM} and Fig.~\ref{fig:supp_JSI_afterDWDM}, we take the latter approach by matching the pump linewidth to the DWDM channel bandwidth and increasing the effective crystal length. As a result, the DWDM-filtered JSI shows a near Gaussian profile, as assumed in Appendix~\ref{app freq conv}.

Lastly, we consider the case of having non-ideal detectors, with the inverse of the jitter time being \textit{smaller} than 12.5~GHz. In this scenario, the two-photon interference visibility is determined by the purity of the DWDM-filtered state.

We follow Ref.~\cite{Ou_1999}'s formalism to compute the two-photon interference visibility $v$ with gating (heralding):
\begin{align}
v &= \frac{\mathcal{E}}{\mathcal{A}},
\end{align}
where
\begin{align}
\mathcal{E} &= (g^{(1)})^2\int d\omega_1 d\omega_1' d\omega_2 d\omega_2' J(\omega_1,\omega_2)J(\omega_1',\omega_2')J^*(\omega_1,\omega_2')J^*(\omega_1',\omega_2)\\
\mathcal{A} &= (g^{(1)})^2\int d\omega_1 d\omega_1' d\omega_2 d\omega_2' |J(\omega_1,\omega_2)J(\omega_1',\omega_2')|^2
\end{align}
We find the theoretical maximum visibility to be $v=0.996$.

Finally, we estimate the entanglement swap fidelity $\mathcal{F}_{\text{swap}}=(1+v)/2$ to be 0.998, which still exceeds 0.99, highlighting the importance of JSA engineering in the case of imperfect detection. However, we note that the spectral profile of the heralded photonic Bell state would be convolved with the detector's instrument response function, likely demanding further shaping on the MC's pump pulse. For this reason, we assume a sufficiently small detector jitter and leave the effect of imperfect detection for future studies.

\section{Frequency and Bandwidth Conversion}
\label{app freq conv}
Recall that the quantum state of the cavity is%
\begin{align}\label{quantum state cav def}
    \ket{\Psi_{\rm{cav}}} \equiv \psi_a(t) \ket{10} +  \psi_b(t) \ket{01}, 
\end{align} 
where $\ket{10}$ corresponds to the photon being in mode $a$ while $\ket{01}$ corresponds to the photon being in mode $b$. The equations of motion for the Schr\"odinger coefficients are 

\begin{subequations}\label{eoms two modes one photon absorption}
\begin{align}
    \dot{\psi}_{a}(t)  &=  -\Big(i\delta_a+\frac{\kappa_a}{2} \Big)\psi_{a}(t) -i|\Lambda(t)|e^{-i\phi(t)}\psi_{b}(t) + \sqrt{\kappa_{a,\rm{w}}}\xi_{a,\rm{i}}(t) \label{EOM cp a}\\
    \dot{\psi}_{b}(t)  &= -\Big(i\delta_b+\frac{\kappa_b}{2}\Big)\psi_{b}(t)  -i|\Lambda(t)|e^{i\phi(t)}\psi_{a}(t) \label{EOM cp b}\\
     \xi_{a,\rm{o}}(t)  &= \xi_{a,\rm{i}}(t) - \sqrt{\kappa_{a,\rm{w}}}\psi_{a}(t) \label{EOM out a}\\
     \xi_{b,\rm{o}}(t)  &= - \sqrt{\kappa_{b,\rm{w}}}\psi_{b}(t) \label{EOM out b},
\end{align} 
\end{subequations}
where the photon wave packets are described by the functions $\xi_{q,\rm{i}}(t)$ and  $\xi_{q,\rm{o}}(t)$ with $q\equal \{a,b\}$. The input state may be expressed in a time-bin basis as~\cite{Heuck_2020_PRA} 
\begin{align}\label{quantum state wg input def}
    \ket{\Psi_{\rm{in}}} \equiv  \int d t \xi_{a,\rm{i}} \hat{w}_a^\dagger(t) \ket{\boldsymbol{0}}_t,
\end{align} 
with $\ket{\boldsymbol{0}}_t$ representing the product state where all time-bins are empty (i.e.\ multi temporal-mode vacuum state) and $\hat{w}_a^\dagger (t)$ populates the time-bin indexed by $t$ with one photon. The output state is 
\begin{align}\label{quantum state wg output def}
    \ket{\Psi_{\rm{out}}} \equiv   \int d t \xi_{a,\rm{o}}(t) \hat{w}_a^\dagger(t) \ket{\boldsymbol{0}}_t +  \int d t \xi_{b,\rm{o}} (t) \hat{w}_b^\dagger(t) \ket{\boldsymbol{0}}_t.
\end{align} 
Note that the control field was divided into amplitude, $|\Lambda(t)|$ , and phase, $\exp[i\phi(t)]$, in Eq.(~\ref{EOM cp a})-(~\ref{EOM cp b}) and the total decay rates are defined by $\kappa_q\equal \kappa_{q,\rm{w}} + \kappa_{q,\rm{l}}$ with $\kappa_{q,\rm{l}}$ being the intrinsic loss rates of the cavity modes. We assume that a maximum conversion efficiency is achieved when $\xi_{a,\rm{o}}(t) \equal 0$, so the task is to determine $|\Lambda(t)|$ and $\phi(t)$ such that this is true. Solving Eq.~\eqref{EOM out a} yields $\psi_a(t)\equal \xi_{a,\rm{i}}(t)/\sqrt{\kappa_{a,\rm{w}}}$, which we substitute into Eq.~\eqref{EOM cp b} and re-arrange terms
\begin{align}\label{EOM cp ai+ 2}
	\frac{d}{dt} \Big(\psi_{b}(t) e^{-Q(t)} \Big)e^{Q(t)}	= \frac{-i}{\sqrt{\kappa_{a,\rm{w}}}} |\Lambda(t)| e^{i\phi(t)} \xi(t) ~\Rightarrow ~~ \psi_{b}(t)	= \frac{-i}{\sqrt{\kappa_{a,\rm{w}}}} e^{Q(t)}  \int_{t_0}^t\! e^{-Q(s)}  |\Lambda(s)|e^{i\phi(s)} \xi(s) ds .
\end{align} 
We defined the function $Q(t)=  -(i\delta_b + \kappa_{b}/2) t$ and replaced $\xi_{a,\rm{i}}$ with $\xi$ for brevity in Eq.~\eqref{EOM cp ai+ 2}. Substituting $\psi_{a} \equal \xi/\sqrt{\kappa_{a,\rm{w}}}$ into Eq.~\eqref{EOM cp a} yields
\begin{align}\label{EOM cp as 2a}
	 \frac{(\kappa_{a,\rm{w}}-\kappa_{a,\rm{l}})}{2}\xi(t) - \dot{\xi}(t) - i\delta_a \xi(t) 	&=  i |\Lambda(t)| e^{-i\phi(t)}\sqrt{\gamma}\psi_{b}(t).
\end{align} 
Multiplying Eq.~\eqref{EOM cp as 2a} by $\xi^{*}(t)\!\exp(\kappa_{b} t)$ and defining real functions $f_i$ and $g_i$, we find
\begin{align}\label{EOM cp as 3}
	f_i(t) + i g_i(t)	&=   |\Lambda(t)|e^{-i\phi(t)}\xi^*(t) e^{(-i\delta_b+\frac{\kappa_{b}}{2})t} \int_{\!t_0}^t\! e^{(i\delta_b+ \frac{\kappa_{b}}{2})s}  |\Lambda(s)|e^{i\phi(s)} \xi(s)ds,
\end{align} 
with 
\begin{subequations}\label{fi gi definition}
\begin{align}
	 f_i(t) &= \Big(\frac{\kappa_{a,\rm{w}}-\kappa_{a,\rm{l}}}{2} \xi(t) - \dot{\xi}(t)\Big)\xi^*(t) e^{\kappa_{b}t} \label{fi definition}\\
     g_i(t) &= -\delta_a |\xi(t)|^2 e^{\kappa_{b}t}. \label{gi definition}
\end{align} 
\end{subequations}
Note that Eq.~\eqref{fi definition} assumes an input wavepacket without chirp, $\frac{d}{dt}[\arg{\xi(t)}]\equal 0$. The right hand side of Eq.~\eqref{EOM cp as 3} can be written as
\begin{align}\label{complex product}
	 \big[x(t)-iy(t)\big] \int_{t_0}^t\! \big[x(s)+iy(s)\big]ds =  x(t)\int_{t_0}^t\!x(s)ds + y(t)\int_{t_0}^t\! y(s)ds + i\Big(x(t) \int_{t_0}^t\! y(s)ds - y(t)\int_{t_0}^t\! x(s)ds\Big),
\end{align} 
where 
\begin{subequations}\label{x and y chi-3}
\begin{align}
	 x(t)   &=   |\Lambda(t)||\xi(t)| \exp(\kappa_{b}t/2) \cos\!\big[\phi(t)+\delta_b t +\arg(\xi)\big]\\
	 y(t)   &=  |\Lambda(t)||\xi(t)| \exp(\kappa_{b}t/2)\sin\!\big[\phi(t)+\delta_b t+\arg(\xi)\big].
\end{align} 
\end{subequations}
By defining the functions
\begin{align}\label{aux functions}
	 X(t)=\int_{t_0}^t\!x(s)ds = R(t)\cos\!\big[\theta(t)\big], ~ Y(t)=\int_{t_0}^t\!y(s)ds = R(t)\sin\!\big[\theta(t)\big],
\end{align} 
Eq.~\eqref{EOM cp as 3} can be split into real and imaginary parts
\begin{align}\label{EOM cp as 4a}
	 f_i = \dot{X}X + \dot{Y}Y, \qquad 	 g_i = \dot{X}Y - \dot{Y}X.
\end{align} 
Using the definition in Eq.~\eqref{aux functions}, we have
\begin{multline}\label{RHS real}
	 f_i = \dot{X}X + \dot{Y}Y = \big[\dot{R}\cos(\theta) - R\sin(\theta)\dot{\theta}\big]R\cos(\theta) + \big[\dot{R}\sin(\theta) + R\cos(\theta)\dot{\theta} \big]R\sin(\theta) = \dot{R}R = \frac12 \frac{d}{dt}\Big( R^2\Big),     
\end{multline} 
which has the solution
\begin{align}\label{R sol}
	 R(t) = \sqrt{ 2\int_{t_0}^t\! f_i(s)ds}.
\end{align} 
Similarly, 
\begin{align}\label{RHS imag}
	 g_i = \dot{X}Y - \dot{Y}X = \big[\dot{R}\cos(\theta) - R\sin(\theta)\dot{\theta}\big]R\sin(\theta) - \big[\dot{R}\sin(\theta) + R\cos(\theta)\dot{\theta} \big]R\cos(\theta) = -R^2\dot{\theta}.
\end{align} 
Using the result in Eq.~\eqref{R sol}, the solution for $\theta$ is 
\begin{align}\label{theta sol}
	 \theta(t) = -\frac12 \int_{t_0}^t\!\! \frac{g_i(s)}{ \int_{t_0}^s\! f_i(z)dz}ds.
\end{align} 
To find the solution for $|\Lambda(t)|$ we evaluate $x^2+y^2\equal |\Lambda|^2 |\xi|^2 \exp(\kappa_{b} t)$ using the results above
\begin{multline}\label{absolute value absorption}
 |\Lambda|^2 |\xi|^2e^{\gamma_L t} =  \dot{X}^2 + \dot{Y}^2 = \big[\dot{R}\cos(\theta) - R\sin(\theta)\dot{\theta}\big]^2 + \big[\dot{R}\sin(\theta) + R\cos(\theta)\dot{\theta} \big]^2 = \dot{R}^2 + R^2\dot{\theta}^2 = \frac{1}{2\int\!f_i} \big( g_i^2 + f_i^2\big) .    
\end{multline} 
Inserting the definition of $g_i$ from Eq.~\eqref{gi definition} yields
\begin{align}\label{p sol} 
	 |\Lambda|^2 |\xi^2| \exp(\kappa_{b} t) = \frac{1}{2\mathcal{F}_i} \Big( \delta_a^2  \exp(2\kappa_{b} t)|\xi|^4 + f_i^2\Big) ~ \Rightarrow ~ 
	 |\Lambda(t)|^2 = \frac{\delta_a^2  |\xi|^2 e^{\kappa_{b} t} }{2\mathcal{F}_i} +   \frac{f_i^2  e^{-\kappa_{b} t} }{2 |\xi(t)|^2 \mathcal{F}_i} ,
\end{align} 
where $\mathcal{F}_i(t)$ is the anti-derivative of $f_i(t)$. If $\delta_a\equal 0$, the solution is
\begin{align}\label{LAM delta_a = 0}
	 |\Lambda(t)| = \frac{|f_i(t)| e^{-\frac{\kappa_{b}}{2}t}}{\sqrt{2} |\xi(t)|}\frac{1}{\sqrt{ \displaystyle \int_{t_0}^{t}  f_i(s) ds } }.
\end{align}
Knowing $|\Lambda(t)|$ means $g_i$ is a known function and $x$ and $y$ may be evaluated using $\theta$ from Eq.~\eqref{theta sol}. Then, the phase $\phi$ is
\begin{align}\label{phi sol}
	 \phi(t) &= -\delta_b t - \arg(\xi) + \tan^{-1}\!\bigg( \frac{y(t)}{x(t)}\bigg) .
\end{align}
%
To obtain $x$ and $y$, note that 
\begin{align} 
	x & = \dot{X} = \dot{R}\cos(\theta) - R\sin(\theta)\dot\theta = \frac{f_i \cos(\theta) + g_i \sin(\theta)}{\sqrt{ 2\int\!f_i}} \\ 
	y & = \dot{Y} = \dot{R}\sin(\theta) + R\cos(\theta)\dot\theta = \frac{f_i \sin(\theta) - g_i \cos(\theta)}{\sqrt{ 2\int\!f_i}}.
\end{align} 
When $\delta_a\equal 0$, we have $g_i\equal 0$ and Eq.~\eqref{phi sol} simplifies to 
\begin{align}\label{phi sol simple}
	 \phi(t) &= -\delta_b t - \arg(\xi) .
\end{align}

\begin{figure}[!h] 
\centering
    \includegraphics[width=\textwidth] {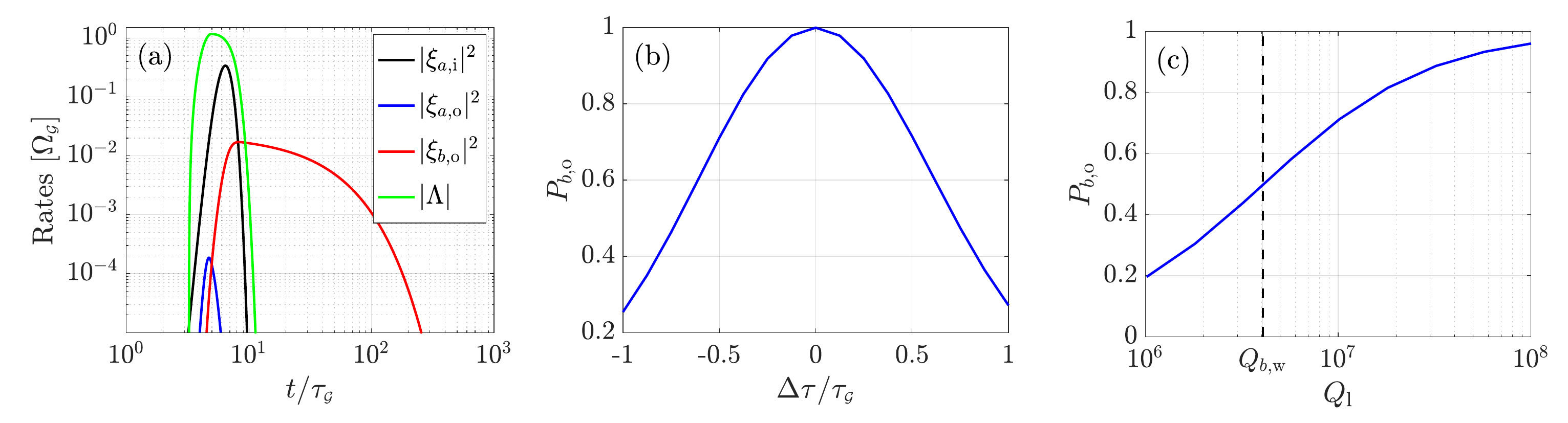}
\caption{Simulation results for frequency conversion and bandwidth narrowing. (a) Example of the solutions to~\eqref{eoms two modes one photon absorption}. (b) Conversion efficiency as a function of the timing-offset between the incident photon and the control field. (c) Conversion efficiency as a function of the intrinsic loss rate of the cavity modes.  Parameters: $\tauG\equal 80\,$ps, $\kappa_{a,\rm{w}}\equal 4\OmG$, $\kappa_{b,\rm{w}}\equal 2\pi\!\times\! 100\,$MHz, and $\kappa_{a,\rm{l}}\equal \kappa_{b,\rm{l}}\equal \kappa_{\rm{l}}$. In (a) and (b) we used $\kappa_{\rm{l}}\equal 0$ and in (a) and (c) we used $\Delta\tau\equal 0$. }
\label{freq conv example}
\end{figure}

Using the same example as the one presented in the main text, let us consider Gaussian input pulses
\begin{align}\label{app Gaussian}
\xi_{a,\rm{i}}(t) &=  \!\sqrt{\frac{2}{\tauG}} \!\left(\frac{\text{ln}(2)}{\pi}\right)^{\!\!\frac{1}{4}} \!\!\exp\!\left(\!-2\text{ln}(2)\frac{(t-T_{\rm{in}})^2}{\tauG^2} \right), 
\end{align} 
where $|\xi_{a,\rm{i}}(t)|^2$ has a full temporal width at half maximum (FWHM) of $\tauG$, spectral width of $\OmG\equal 4\text{ln}(2)/\tauG$, and integrates to 1 (over the infinite interval from $-\infty$ to $\infty$). We also introduce the offset $\Delta\tau$ to investigate the influence of timing-offset between the incident photon and the control pulse. The control pulse, $\Lambda(t)$, is then calculated using $\xi(t+\Delta\tau)$ instead of $\xi(t)$. Solving the equations of motion in Eq.~\eqref{eoms two modes one photon absorption} with the control field calculated from Eq.~\eqref{LAM delta_a = 0}--\eqref{phi sol simple}, we calculate the probability of converting the photon to $\omega_b$ as 
\begin{align}\label{app p freq conv}
P_{b,\rm{o}} &= \int | \xi_{b,\rm{o}}(t)|^2 dt .
\end{align} 
The results are shown in Fig.~\ref{freq conv example}.
We use an input pulse centered at 1550$\,$nm with a temporal width of $\tauG\equal 80\,$ps and $\kappa_{a,\rm{w}}\equal 4\OmG$, which ensures efficient absorption into cavity mode $a$. Fig.~\ref{freq conv example}(a) plots the input and output wave packets for an example with $\kappa_{a,\rm{l}}\equal \kappa_{b,\rm{l}}\equal \kappa_{\rm{l}}\equal  0$, $\Delta\tau\equal 0$, and $\kappa_{b,\rm{w}}\equal 2\pi\!\times\! 100\,$MHz to match the bandwidth of the output pulse to the cavity containing the spin qubit. The small amplitude of the blue line in Fig.~\ref{freq conv example}(a) shows that a very small fraction of the incident wave packet passes by the cavity without being absorbed. This fraction decreases towards zero very rapidly as $\kappa_{a,\rm{w}}$ is increased relative to $\OmG$~\cite{Heuck_2020_PRA}. The effect of timing mismatch is investigated in Fig.~\ref{freq conv example}(b), which shows that the conversion efficiency stays above 70\% if $|\Delta\tau|\leq \tauG$. Note that more robustness could be obtained by optimizing the control pulse while taking the timing-offset into account. Fig.~\ref{freq conv example}(c) shows the reduction in conversion efficiency in the case of a finite intrinsic decay rate of the cavity modes (assumed to be equal for modes $a$ and $b$). The required narrow bandwidth of the output pulse requires the loaded $Q$ to be larger than $Q_{b,\rm{w}}\equal 4\!\times\!10^6$. When $Q_{\rm{l}}\equal Q_{b,\rm{w}}$, the spectrum of the output pulse is roughly Lorentzian with a FWHM bandwidth of 200$\,$MHz and the conversion efficiency is 50\% as illustrated using the vertical dased black line in Fig.~\ref{freq conv example}(c).

\section{Switching array in the receiver} \label{app_switch_array}
\begin{figure*}[t]
    \centering
    \includegraphics[width=0.5\textwidth]{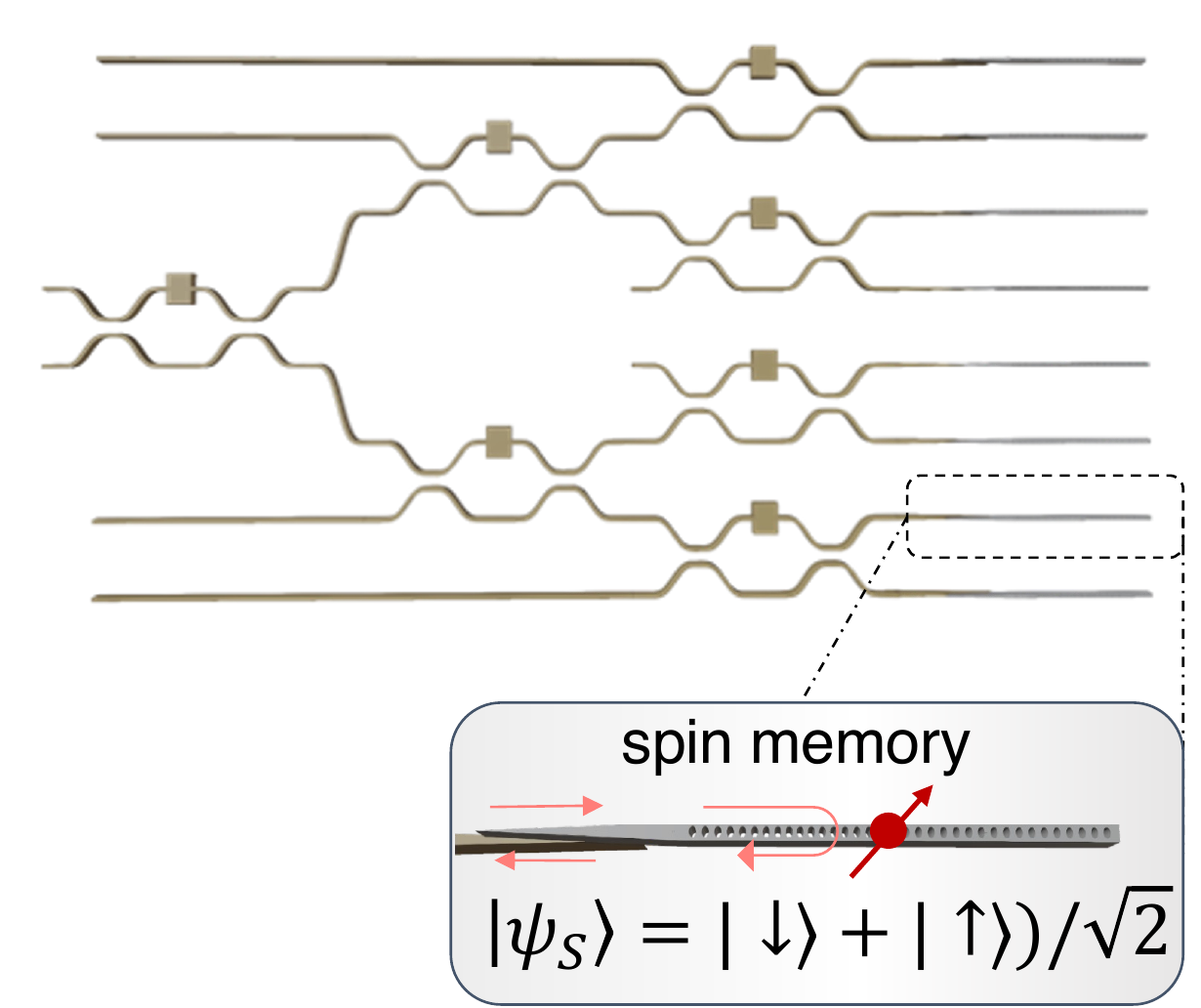}
    \caption{An illustration of a MZI tree network as the interposer to the memories. Each layer introduces transmission loss from the MZI.}
    \label{fig:supp_MZI_tree}
\end{figure*}

\begin{figure*}[t]
    \centering
    \includegraphics[width=0.5\textwidth]{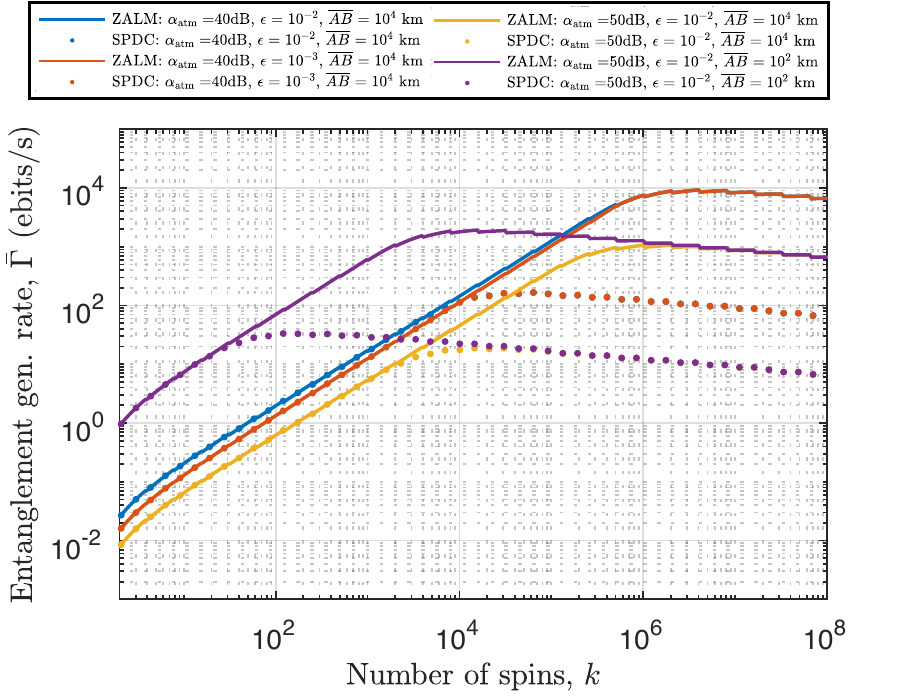}
    \caption{Entanglement generation rate $\bar{\Gamma}$ vs $k$ for both ZALM and SPDC with varying $\epsilon\in\{10^{-2},10^{-3}\}$ and total downlink atmospheric attenuation $\alpha=\{40,50\}~$dB. We additionally consider $\overline{AB}=10^2~$km (purple). The rates are based on a scheme that employs a MZI tree network that has $k$-dependent transmission loss.}
    \label{fig:supp_MZI_tree_rate}
\end{figure*}

Here we consider using a MZI tree network in a photonic integrated circuit (PIC) as a fast switching array, as shown in Fig.~\ref{fig:supp_MZI_tree}, albeit with finite transmission loss that scales exponentially with the number of layers. For LiNbO$_3$-based PIC, propagation loss has been shown to be negligible even for device length exceeding $100~\mu$m~\cite{Zhang_2017}. For simplicity, we neglect metal absorption loss stemming from nearby electrodes used to drive the electro-optic phase shifters. Rather, we assume the main loss mechanism to come from imperfection in the directional couplers. We take a state-of-the-art value of $\sim$0.2~dB per MZI demonstrated on SOI~\cite{Wilmart_2021} to illustrate the effect of switching array loss on the entanglement generation rate. Fig.~\ref{fig:supp_MZI_tree_rate} shows comparable rates as the non-MZI-based scheme shown in the main text, up until $k\approx 10^4$. After which point, the $k$-dependent switching array loss begins reducing the entanglement generation rate.

This particular qRX setup requires heterogeneous integration of diamond color centers into PICs, a feat which has already been demonstrated by Ref.~\cite{Wan_2020} and is conducive to scaling up a multiplexed quantum repeater network. Despite solid-state emitters such as SiV$^-$ manifesting spectral inhomogeneity, we argue post-selecting candidates within a narrowed inhomogeneous distribution and performing subsequent in-situ tuning enabled by an active PIC platform could still ensure maximal spin-cavity coupling. For example, SiV$^-$ can be strain-tuned~\cite{Meesala_2018} to shift its optical transition frequency, while the nanophotonic cavity's resonance can be gas-tuned~\cite{Faraon_2012} (i.e. index shifting).

\section{Spin-photon system} \label{app_spin_cavity}
\subsection{Silicon-vacancy center}
The proposed architecture considers diamond's negatively-charged silicon-vacancy center as the atomic memory. With an applied magnetic field and accounting for only Zeeman splitting, the energy ground state splits into two electron spin states $\ket{\downarrow}$ and $\ket{\uparrow}$. One of the two optical transitions with the excited state $\ket{e}$, $\ket{\downarrow}\longleftrightarrow\ket{e}$, is coupled with the cavity mode. With a nearby nuclear spin, hyperfine splitting further divides the two electronic spin states to a total of four levels: $\ket{\downarrow_e\downarrow_n},\ket{\downarrow_e\uparrow_n},\ket{\uparrow_e\uparrow_n},\ket{\uparrow_e\downarrow_n}$. Effectively, the electron spin acts as a broker qubit that interfaces with the photon, and subsequently transfers the qubit state to the nuclear spin that serves as a long-lived atomic memory~\cite{Nguyen_2019_PRB}.
\begin{figure*}[t]
    \centering
    \includegraphics[width=0.5\textwidth]{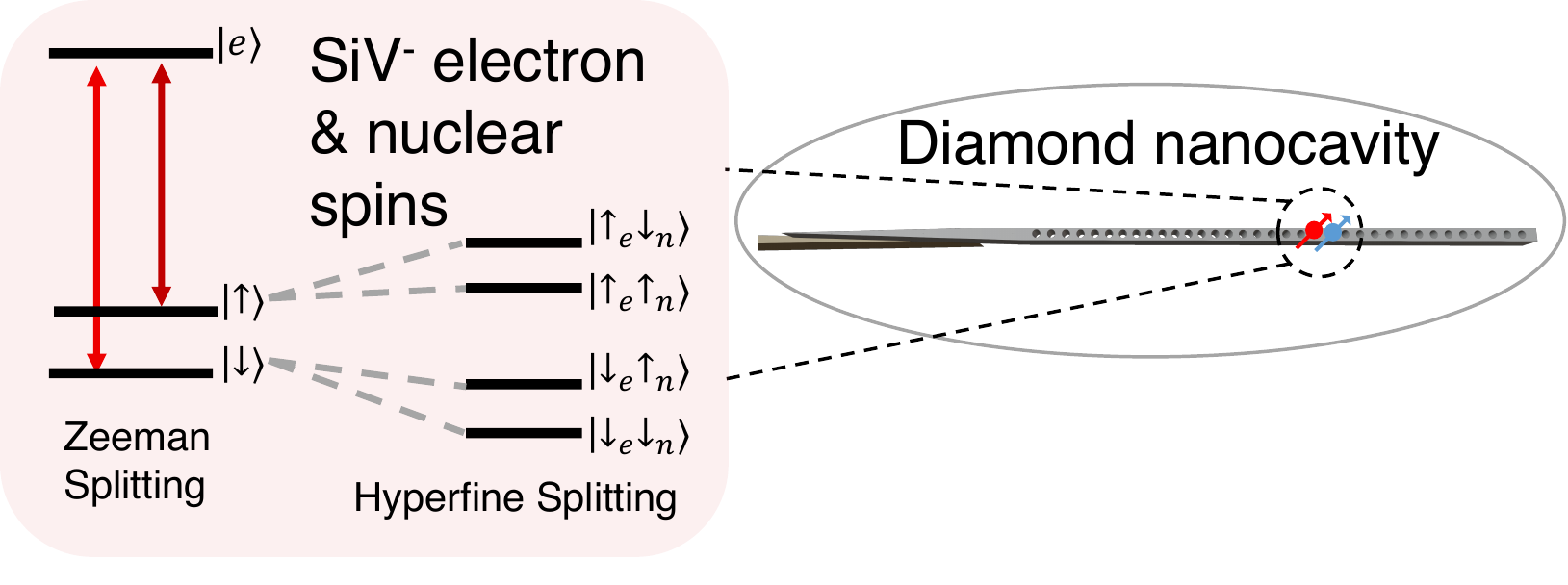}
    \caption{Level-structure of a SiV$^-$ center in diamond. The electron spin (red) contains two Zeeman-split states $\{\ket{\downarrow},\ket{\uparrow}\}$, which further split into four states $\{\ket{\downarrow_e\downarrow_n},\ket{\downarrow_e\uparrow_n},\ket{\uparrow_e\uparrow_n},\ket{\uparrow_e\downarrow_n}\}$ due to hyperfine coupling with the nuclear spin (blue).}
    \label{fig:siv_level_structure}
\end{figure*}

\subsection{Spin-dependent cavity reflection}
The reflection coefficient of a single-sided cavity coupled with a quantum emitter is:
\begin{align}\label{eqn:supp:cav_reflection}
r(\omega) &= 1-\frac{\kappa_{\text{wg}}\left(i\Delta_a+\frac{\gamma}{2}\right)}{\left(i\Delta_c+\frac{\kappa}{2}\right)\left(i\Delta_c+\frac{\gamma}{2}\right)+g^2}
\end{align}
where $g$ is the atom-cavity coupling strength, $\gamma$ is the emitter's spontaneous emission rate, $\kappa$ is the cavity's total decay rate, $\kappa_{\text{wg}}$ is the waveguide-cavity coupling rate, and $\Delta_a=\omega_a-\omega$ and $\Delta_c=\omega_c-\omega$ are the atomic and cavity detuning from the probe photon, respectively. In the large cooperativity $C=4g^2/\kappa\gamma\gg 1$ limit and on resonance $\Delta_a=\Delta_c=0$, the reflection coefficient of a perfectly over-coupled cavity simplifies to
\begin{align}
r(\omega) \xrightarrow{C\gg 1} \frac{C-1}{C+1}
\end{align}
Therefore, $r$ approaches $+1$ when $C$ increases, whereas a emitter decoupled from the cavity mode would yield $r\rightarrow-1$. We consider a spin qubit whose basis states are $\ket{\downarrow}$ and $\ket{\uparrow}$. When the emitter is in the $\ket{\downarrow}$ state that is coupled to the cavity mode, $r=+1$. On the other hand, if it is in the $\ket{\uparrow}$ state that is decoupled with the cavity mode, $r=-1$. As a result of this state-dependent phase difference, the probe photon is entangled with the spin via cavity reflection.

\subsection{Photon-to-spin mapping}
In the Schrodinger picture, we present an example of how an arbitrary photonic qubit encoded in the polarization basis $\{\ket{H},\ket{V}\}$, ${\ket{\psi}_P=\alpha\ket{H}+\beta\ket{V}}$ ($\equiv\alpha\ket{0,1}+\beta\ket{1,0}$ in the dual-rail Fock basis), can be teleported to a spin qubit ${\ket{\psi}_{S,\text{final}}=\alpha\ket{\downarrow}+\beta\ket{\uparrow}}$. We initialize the spin to be in an equal superposition state: ${\ket{\psi}_{S,\text{init}}=\left(\ket{\downarrow}+\ket{\uparrow}\right)/\sqrt{2}}$. Their joint state is then: $\ket{\Psi}=\ket{\psi}_P\otimes\ket{\psi}_{S,\text{init}}={\alpha\ket{H,\downarrow}+\alpha\ket{H,\uparrow}+\beta\ket{V,\downarrow}+\beta\ket{V,\uparrow}}$. Upon entering the receiver node, we envision using a polarization-splitter rotator to convert the polarization basis to the spatial basis $\{a_H,a_V\}$. Importantly, we re-write the joint state as ${\ket{\Psi}=\alpha\ket{a_H,\downarrow}+\alpha\ket{a_H,\uparrow}+\beta\ket{a_V,\downarrow}+\beta\ket{a_V,\uparrow}}$. Subsequently, mode $a_H$ acquires a spin-dependent phase upon cavity reflection, whereas mode $a_V$ acquires a constant $-1$ phase from reflection off a mirror. The resultant state is
\begin{align}
\ket{\Psi} &= \alpha\ket{a,\downarrow}-\alpha\ket{a,\uparrow}-\beta\ket{b,\downarrow}-\beta\ket{b,\uparrow}
\end{align}

We note describe the evolution of each of the two spatial modes separately. After cavity interaction, $\{\ket{a_H},\ket{a_V}\}$ enter a 50:50 beam splitter whose output modes are $\{\ket{A}=t\ket{a_H}+r\ket{a_V},\ket{B}=r\ket{a_H}+t\ket{a_V}\}$, where $r=i$ and $t=1$ are the reflection and transmission coefficients, respectively. The two output modes then become
\begin{align}
\ket{A}&=\alpha\left(\ket{\downarrow}-\ket{\uparrow}\right)-i\beta\left(\ket{\downarrow}+\ket{\uparrow}\right)\\
\ket{B}&=i\alpha\left(\ket{\downarrow}-\ket{\uparrow}\right)-\beta\left(\ket{\downarrow}+\ket{\uparrow}\right)
\end{align}

The spin undergoes a Hadamard rotation, transforming the states to 
\begin{align}
\ket{A}&=\alpha\ket{\uparrow}-i\beta\ket{\downarrow}\\
\ket{B}&=i\alpha\ket{\uparrow}-\beta\ket{\downarrow}
\end{align}

Upon detection on either of the $\ket{A}$ or $\ket{B}$ port, appropriate Pauli operations can be applied to the spin qubit to obtain the target state ${\ket{\psi}_{S,\text{final}}=\alpha\ket{\downarrow}+\beta\ket{\uparrow}}$.

However, with imperfections such as finite cooperativity and non-unity coupling to the waveguide mode ($\kappa_{\text{wg}}/\kappa<1$) in the spin-cavity system, the reflection coefficients would have non-unity amplitudes. We consider now the photon-to-spin mapping process with generalized reflection coefficients, $\ron,\roff$, for the on- and off-resonance cases respectively. We still treat the mirror as a lossless component such that mode $b$ still acquires a constant -1 phase.

Upon reflection, the spin-photon state is
\begin{align}
\ket{\Psi} &= \ron\alpha\ket{a_H,\downarrow}+\roff\alpha\ket{a_H,\uparrow}-\beta\ket{a_V,\downarrow}-\beta\ket{a_V,\uparrow}
\end{align}

After the 50:50 beam splitter, the two output modes are
\begin{align}
\ket{A}&=\alpha\left(\ron\ket{\downarrow}+\roff\ket{\uparrow}\right)-i\beta\left(\ket{\downarrow}+\ket{\uparrow}\right)\\
\ket{B}&=i\alpha\left(\ron\ket{\downarrow}+\roff\ket{\uparrow}\right)-\beta\left(\ket{\downarrow}+\ket{\uparrow}\right)
\end{align}

\section{Dependence of $\bar{\Gamma}$ on $\eta$} 
\label{app_sqrt_eta}
\subsection{Memory multiplexing dependent rate scaling behavior}
From the main text, the entanglement generation rate is defined to be
\begin{align}
\bar{\Gamma} &= \frac{p_{\text{success}}\cdot (k-1)}{\tau_{\text{idle}}}
\end{align}
where
\begin{align}
p_{\text{success}} &= \eta \left(\frac{1-(1-\sqrt{\eta})^{2N}}{1-(1-\sqrt{\eta})^2}\right)
\end{align}

For simplicity, let us neglect the contribution of $\epsilon$ and define $N=N_k=\tau_{\text{idle}}/((k-1)\tau_0)$. In the memory-limited regime, $\tau_{\text{idle}}/\tau_0\gg (k-1)$, thus $N_k\gg 1$. We can re-write $p_{\text{success}}$ as
\begin{align}
p_{\text{success}} &\approx \eta \left(\frac{1-e^{-2N_k\sqrt{\eta}}}{2\sqrt{\eta}}\right)\\
&\approx \frac{1}{2}\sqrt{\eta}
\end{align}
where we used the approximations $(1+x)^{\alpha}\approx e^{\alpha x}$ for large $|\alpha x|\gg 1$ and $(1+x)^{\alpha}\approx 1+\alpha x$ for small $|\alpha x|\ll 1$. From above, we see that $\bar{\Gamma}$ scales as $\sqrt{\eta}$ in the memory-limited regime.

On the other hand, with sufficiently high memory multiplexing such that $N_k\ll 1$,
\begin{align}
p_{\text{success}} &\approx \eta \left(\frac{2N_k\sqrt{\eta}}{2\sqrt{\eta}}\right)\\
&= N_k\eta
\end{align}
which recovers the typical $\bar{\Gamma}\propto\eta$ scaling.

\section{Spin-spin Bell state generation rate fidelity trade-off} \label{app_rate_fidelity}

\begin{figure}
    \centering
    \includegraphics[width=0.5\textwidth]{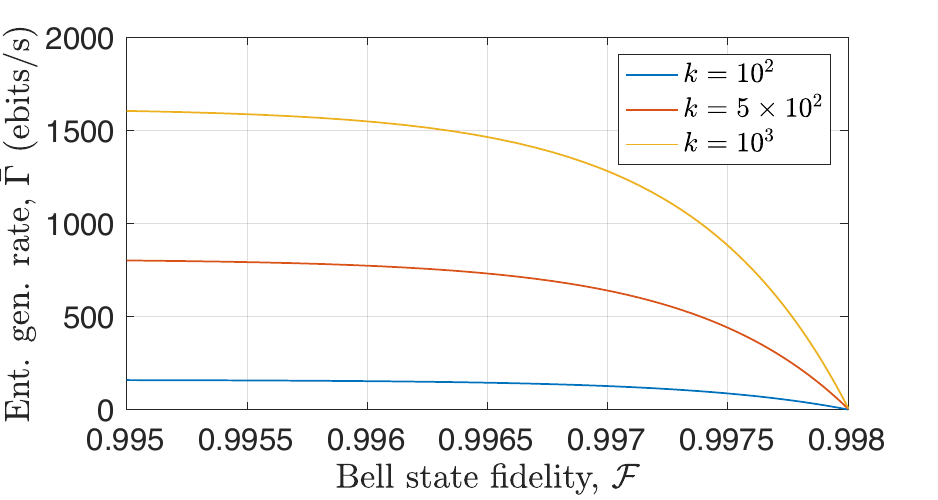}
    \caption{Rate-fidelity trade-off at ${k\in\{10^2,5\times 10^2,10^3\}}$. Setting $N_s=2.94\times 10^{-2}$ and $C=100$, $\mathcal{F}$ upper-bounds at 0.998 due to imperfections in the qTX and qRX. We take $\overline{AB}=10^2$~km and $\alpha_\text{atm}=40$~dB.}
    \label{fig:app_rate_fidelity_tradeoff}
\end{figure}

Figure~\ref{fig:app_rate_fidelity_tradeoff} demonstrates the rate-fidelity trade-off at different ${k\in\{10^2,5\times 10^2,10^3\}}$. $\bar{\Gamma}$ approaches zero as $\mathcal{F}$ increases. Similar to what is shown in Fig.~\ref{fig:k_rate} with $\overline{AB}=10^2$~km and $\alpha_\text{atm}=40$~dB, $\bar{\Gamma}$ increases monotonically with the number of spins.

\section{ZALM for ground-only quantum networks} \label{app_ground_networks}

\begin{figure}
    \centering
    \includegraphics[width=0.75\textwidth]{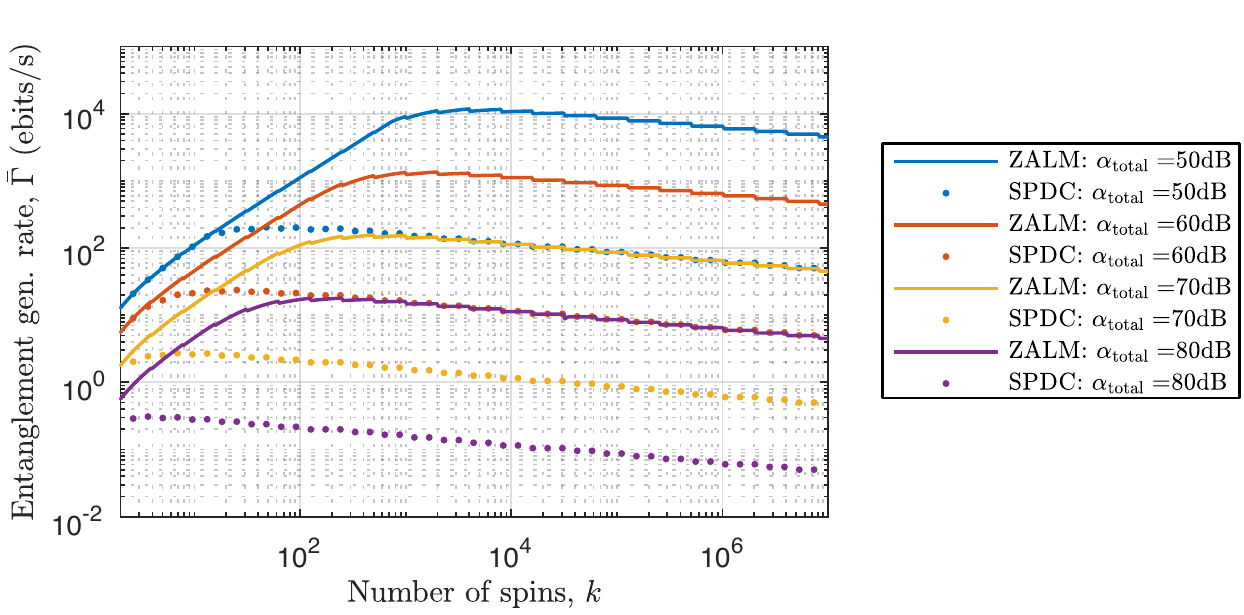}
    \caption{Entanglement generation rate $\bar{\Gamma}$ as a function of the number of spins $k$. $\bar{\Gamma}$ vs $k$ for both ZALM and SPDC with varying channel losses $\alpha=\{50,60,70,80\}~$dB. These calculations assume $\overline{AB}=10^2$~km with corresponding $\tau_\text{comm}=30$~ms and $\epsilon=10^{-3}$.}
    \label{fig:supp_rate_ground_only}
\end{figure}

The quasi-deterministic ZALM BPS is useful for general two-way quantum repeater networks regardless of their configurations. Although the main text provides a specific example for a satellite-assisted architecture for global scale networks, we show here that the same qTX can be equally beneficial for ground-only quantum networks. Figure~\ref{fig:supp_rate_ground_only} shows the rate of generating entanglement between A and B, with a midpoint source C equidistant from the two qRX's. We assume $\overline{AB}=2L_\text{GG}=10^2$~km with corresponding classical communication time of 30~ms and $\epsilon=10^{-3}$. Again, we compare the rate performance between ZALM and a free-running narrowband-filtered SPDC for channel losses $\alpha=\{50,60,70,80\}~$dB accounting for the attenuation loss in optical fibers. Additionally, we assume the qRX containing a fast-switching PIC whose MZI tree array causes compounded loss dependent on the number of tree layers. In the memory-limited regime with $k\geq 10$, ZALM already outperforms SPDC by at least an order of magnitude in $\bar{\Gamma}$. The gap in rate widens further with greater channel loss. For example, the difference in $\bar{\Gamma}$ at $k=1$ is much larger for $\alpha=80$~dB than for $\alpha=50$~dB. Nevertheless, with increasing memory multiplexing, the rate advantage in using the ZALM BPS immediately manifests.

\twocolumngrid

\bibliography{references_satellite}

\end{document}